\documentclass[amssymb,twocolumn,prb,notitlepage,superscriptaddress,floats,amsmath]{revtex4-2}

\usepackage{epsfig}
\usepackage{bm}
\usepackage{color}
\usepackage{array}

\begin{document}

\title{%Exciton spectra in WSe$_2$ multilayers
Rydberg series of intralayer K-excitons in WSe$_2$ multilayers}

\author{Piotr Kapuscinski}
 \email{piotr.kapuscinski@fuw.edu.pl}
 \affiliation{Laboratoire National des Champs Magn\'etiques Intenses, LNCMI-EMFL,
CNRS UPR3228,Univ. Grenoble Alpes, Univ. Toulouse,
Univ. Toulouse 3, INSA-T, Grenoble and Toulouse, France}
 \affiliation{Institute of Experimental Physics, Faculty of Physics,
University of Warsaw, ul. Pasteura 5, PL-02-093 Warsaw, Poland}
\author{Artur O. Slobodeniuk}
 \affiliation{Department of Condensed Matter Physics, Faculty of Mathematics and Physics, Charles University,
Ke Karlovu 5, CZ-121 16 Prague, Czech Republic
}
\author{Alex Delhomme}
 \affiliation{Laboratoire National des Champs Magn\'etiques Intenses, LNCMI-EMFL,
CNRS UPR3228,Univ. Grenoble Alpes, Univ. Toulouse,
Univ. Toulouse 3, INSA-T, Grenoble and Toulouse, France}
 \affiliation{Walter Schottky Institut and TUM School of Natural Sciences,
Technische Universit{\"a}t M{\"u}nchen, Am Coulombwall 4, 85748 Garching, Germany}
\author{Clément~Faugeras}
 \affiliation{Laboratoire National des Champs Magn\'etiques Intenses, LNCMI-EMFL,
CNRS UPR3228,Univ. Grenoble Alpes, Univ. Toulouse,
Univ. Toulouse 3, INSA-T, Grenoble and Toulouse, France}
\author{Magdalena~Grzeszczyk}
 \affiliation{Institute for Functional Intelligent Materials, National University of Singapore, Singapore 117544, Singapore}
\author{Karol~Nogajewski}
 \affiliation{Institute of Experimental Physics, Faculty of Physics,
University of Warsaw, ul. Pasteura 5, PL-02-093 Warsaw, Poland}
\author{Kenji~Watanabe}
 \affiliation{Research Center for Electronic and Optical Materials, National Institute for Materials Science, 1-1 Namiki, Tsukuba 305-0044, Japan}
\author{Takashi Taniguchi}
 \affiliation{Research Center for Materials Nanoarchitectonics, National Institute for Materials Science,  1-1 Namiki, Tsukuba 305-0044, Japan}
\author{Marek Potemski}
 \email{marek.potemski@lncmi.cnrs.fr}
 \affiliation{Laboratoire National des Champs Magn\'etiques Intenses, LNCMI-EMFL,
CNRS UPR3228,Univ. Grenoble Alpes, Univ. Toulouse,
Univ. Toulouse 3, INSA-T, Grenoble and Toulouse, France}
 \affiliation{CEZAMAT, CENTERA Labs, Warsaw University of Technology, PL-02-822 Warsaw, Poland}

\date{\today}

\begin{abstract}
Semiconducting transition metal dichalcogenides of group VI are well-known for their prominent excitonic effects and the transition from an indirect to a direct band gap when reduced to monolayers. While considerable efforts have elucidated the Rydberg series of excitons in monolayers, understanding their properties in multilayers remains incomplete. In these structures, despite an indirect band gap, momentum-direct excitons largely shape the optical response.
In this work, we combine magneto-reflectance experiments with theoretical modeling based on the {\bf k}$\cdot${\bf p} approach to investigate the origin of excitonic resonances in WSe$_2$ bi-, tri-, and quadlayers. For all investigated thicknesses, we observe a series of excitonic resonances in the reflectance spectra, initiated by a ground state with an amplitude comparable to the ground state of the 1$s$ exciton in the monolayer. Higher energy states exhibit a decrease in intensity with increasing energy, as expected for the excited states of the Rydberg series, although a significant increase in the diamagnetic shift is missing in tri- and quadlayers.  
By comparing the experimental observations with theoretical predictions, we discover that the excitonic resonances observed in trilayers originate from two Rydberg series, while quadlayers exhibit four such series, and bilayers host a single Rydberg series similar to that found in monolayers.
\end{abstract}
\maketitle

\section{Introduction}

The recent advancements in the studies of group VI semiconducting transition metal dichalcogenide (S-TMD) monolayers, renowned for their prominent excitonic effects, have significantly improved our understanding of their electronic and optical properties \cite{Mak2016, KoperskiNanophotonics, Pei2019}. To a great extent, this progress can be attributed to extensive research on excitons and their Rydberg series in these monolayers\cite{Molas_2017, Zhang2017, Wang2017, Stier2018, Molas2019, Goryca2019, Lu_2020, Robert2020, Kapuscinski2021, RobertPRL2021, Ren2023}. A similar level of understanding has not yet been achieved for S-TMD multilayers. In contrast to monolayers, the S-TMDs multilayers are indirect bandgap semiconductors\cite{Splendiani2010, Mak2010, Zhao2013}. Nevertheless, their optical response remains governed by excitonic resonances associated with higher energy direct in \textbf{k}-space transitions\cite{Arora2015, C5NR06782K, Molas2017}. Studies of excitons in S-TMD multilayers offer relevant insights into the evolution of optical and electronic properties of these structures upon increasing the number of stacked layers\cite{Arora2017, Arora2019,Slobodeniuk2019, Zeng2013, Kipczak_2023}. 

The S-TMD multilayers, which can be thought of as stacks of weakly coupled monolayers, form significantly more complex systems where the exciton Rydberg series split into multiple components, with numerous additional yet unknown parameters affecting their properties. Purely theoretical estimation of these parameters does not align well with recent experimental findings. Further works combining the experimental and theoretical efforts are needed to clarify the reported discrepancies. On the experimental side, the observation of excitonic series could provide deeper insights into the electronic properties of multilayers, especially through optical spectroscopy in large magnetic fields, which has historically been a valuable tool for identifying excitonic states. So far, the studies of excitonic resonances in S-TMDs remain rather limited.
Although the ground states of direct and indirect excitons have been studied in several experimental works\cite{Arora2015, C5NR06782K, Molas2017, Arora2017, Arora2019, Slobodeniuk2019, Kipczak_2023}, the full spectra of excitonic resonances, including the ground and excited states, remain largely unexplored.

In this work, we combine magneto-reflectance experiments with theoretical modeling using the {\bf k}$\cdot${\bf p} approach to investigate the origin of excitonic resonances in WSe$_2$ bilayers, trilayers, and quadlayers. Our reflectance spectra reveal a series of excitonic resonances, initiated by a ground state whose amplitude is comparable to that of the 1s exciton ground state in monolayers. By analyzing the diamagnetic shifts of these resonances, we identify the excitonic Rydberg series index  $n$ for the observed transitions.  Comparing our experimental data with theoretical predictions allows us to determine the origins of these resonances, providing insights into the fine structure of excitons in WSe$_2$ multilayers.

\section{Experimental results}

The samples employed in our study consist of a monolayer or multilayer WSe$_2$ encapsulated within hBN flakes and deposited onto a Si substrate or DBR mirror. The heterostructures were fabricated through mechanical exfoliation of bulk WSe$_2$ and hBN crystals. The initial hBN layer was formed via non-deterministic exfoliation onto the Si substrate, while the WSe$_2$ monolayer and the capping hBN flake were transferred using a dry stamping technique with PDMS stamps onto the base hBN flake.

The low-temperature magneto-reflectance experiments were conducted using the Faraday configuration, wherein a magnetic field was applied perpendicular to the plane of the WSe$_2$ layers. This was achieved by employing a free-beam insert within a resistive solenoid capable of generating magnetic fields of up to 30 T. The sample was positioned on top of an x-y-z piezo stage and immersed in gaseous helium at $T=5~$K to maintain the low-temperature conditions.
A tungsten halogen lamp served as the white light source for the experiments. The excitation beam was precisely focused, and the resulting signal was collected using the same microscope objective, featuring a numerical aperture (NA) of 0.83. Circular polarization analysis of the signal was performed using a combination of a quarter wave plate and a linear polarizer. 
Subsequently, the collected light underwent analysis using a 0.5 m long spectrometer equipped with a 500 lines/mm grating and a nitrogen-cooled CCD camera. To improve data clarity, the reflectivity spectra are presented as reflectance contrast (RC), defined as $RC(E)=100\%(R(E)-R_0(E))/(R(E)+R_0(E))$, where $R(E)$ and $R_0(E)$ are the spectra of white light reflected from the structure with and without the WSe$_2$ layer, respectively.

\begin{figure}[t]
	\centering
	\includegraphics[width=\linewidth]{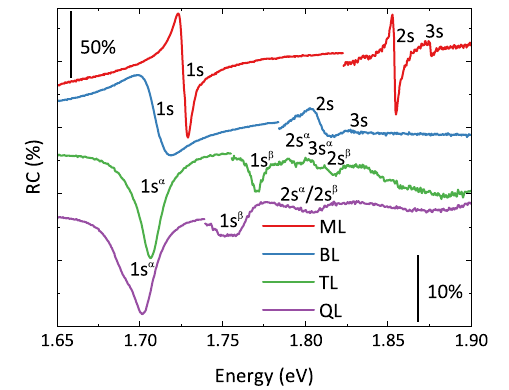}
	\caption{Low temperature ($T=5$~K) RC spectra of WSe$_2$ mono- (red), bi- (blue), tri- (green), and quadlayers (purple) encapsulated in hBN. The spectra were multiplied by a factor of 5 in the spectral region above the excitonic ground state transition. The scale bar on the left indicates the vertical scale for the energy region of the ground states, while the scale bar on the right indicates the scale for the energy region of excited states. The spectra of multilayers were shifted vertically for clarity.}
	\label{fig:exp_intro}
\end{figure}

The RC spectra at low temperature (5~K) of WSe$_2$ mono- (ML), bi- (BL), tri- (TL), and quadlayers (QL) encapsulated in hBN are illustrated in Fig.\ref{fig:exp_intro}. Across all thicknesses, the dominant feature in the spectra is the pronounced resonance at the lowest energy, corresponding to the excitation of the ground intralayer exciton state. Previous investigations have explored the thickness dependence of this particular resonance~\cite{Arora2015}. Following are higher-energy transitions, indicative of excited exciton states. 

\begin{figure*}[t]
	\centering
	\includegraphics[width=\linewidth]{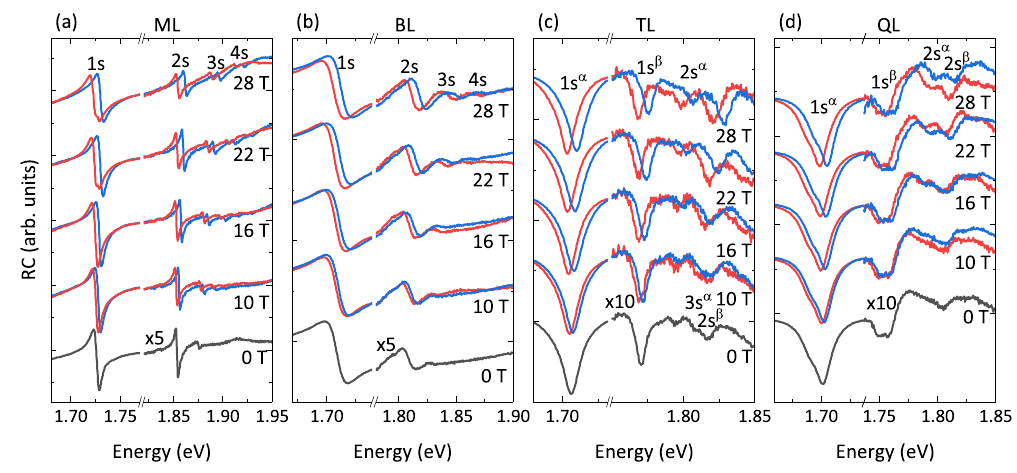}
	\caption{Low-temperature helicity-resolved ($\sigma^+$ - red, $\sigma^-$ - blue lines) RC spectra of (a)~ML, (b)~BL, (c)~TL and~(d) QL at selected magnetic fields $B$ between $0-28$~T. The spectra were multiplied by a factor of 5 in panels (a) and (b) and 10 in (c) and (d) in the spectral region above the excitonic ground state transition.}
	\label{fig:exp_field}
\end{figure*}

The sequence of excitonic states in monolayers is well understood, largely thanks to prior experimental studies~\cite{Stier2018, Molas2019, Liu2019, Chen2019, Wang2020}. These studies have characterized the $s$-type Rydberg series, which can be described either numerically using nonhydrogenic Rytova-Keldysh potential~\cite{Stier2018} or analytically using a modified Kratzer approach~\cite{Molas2019}. In the case of monolayers, we assign the observed resonances to 1$s$, 2$s$, and 3$s$ exciton states based on these findings. The same description, however, cannot be readily extended to thicker layers. The amplitude of transitions related to $ns$ Rydberg states is expected to decrease monotonically with increasing $n(=1,2,3 \dots)$, correlating with higher transition energies. However, this trend does not hold in TL if the observed resonances are treated as a single series. Additionally, the resonances in QL appear to be doubled. These observations hint at a more intricate structure of excitonic series in thicker flakes, consistent with our earlier theoretical predictions~\cite{Slobodeniuk2019}. Specifically, our theoretical framework suggests that optically active transitions in $N$L multilayers involve $N$ series of intensive (so-called intralayer) optical transitions. 

Due to the complexity introduced by this multitude of excitonic resonances in multilayer spectra, low-temperature experiments alone may not suffice for accurately assigning the observed transitions. However, applying a magnetic field offers a way to gain additional insights. By analyzing parameters such as the diamagnetic constant $\alpha$ (which describes the rate of quadratic shift of the transition energy in a magnetic field) as well as the sign and magnitude of the g-factor $g$ (which accounts for the spin-splitting of transitions in a magnetic field), we can correctly attribute the origins of the observed transitions.

We carried out circular polarization-resolved magneto-RC measurements on ML, BL, TL, and QL WSe$_2$ at low temperature, applying magnetic fields up to 30~T perpendicular to the layer plane. In Fig.~\ref{fig:exp_field}, we present the spectra obtained at selected magnetic field strengths. The application of the magnetic field induces two main effects: (i) the lifting of spin degeneracy in excitonic transitions, resulting in the splitting of circularly polarized states, and (ii) a diamagnetic energy shift of all transitions towards higher energies. Notably, for ML and BL, the magnetic field reveals additional features in the spectra, which we attribute to higher excited states.

\begin{figure*}[t]
	\centering
	\includegraphics[width=17.6cm]{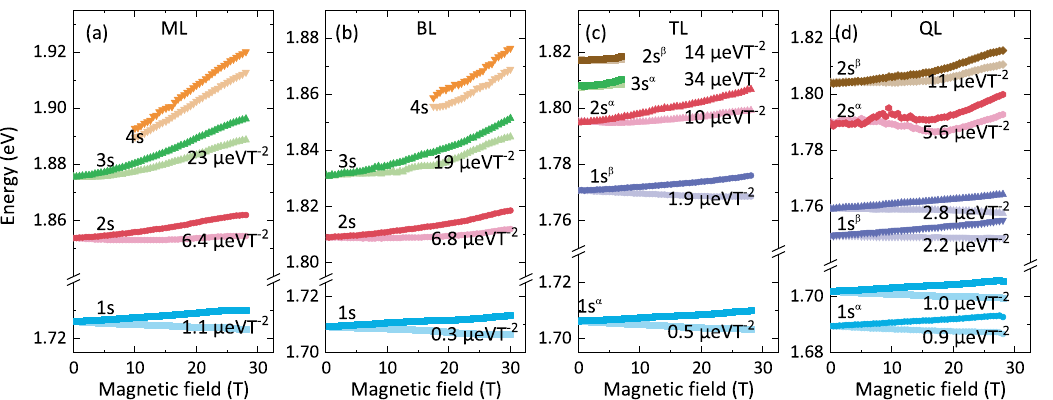}
	\caption{Energies of the circularly polarized $\sigma^+$ (light color) and $\sigma^-$ (dark color)  transitions of the exciton resonances derived from the RC spectra of (a)~ML, (b)~BL, (c)~TL and~(d) QL.}
	\label{fig:fig_energies}
\end{figure*}

To quantify the change in exciton transition energies induced by the applied magnetic field, we fit the Fano curve for ML and BL and the Lorentzian function for TL and QL to the observed resonances. Two different curves were used because the wavelength-dependent phase shift arising from interferences affects the RC excitonic lineshape of the two heterostructure stacks (hBN/WSe$_2$/hBN/Si and hBN/WSe$_2$/hBN/DBR) differently. The resulting energies are depicted in Figure~\ref{fig:fig_energies}. Notably, we observed that all transitions exhibit spin splitting with the same sign and approximately the same magnitude ($g \approx -4$, see Supplementary Material - SM~\cite{Supplementary_Materials}), confirming their association with intralayer exciton transitions~\cite{Arora2017, Slobodeniuk2019}. Furthermore, we extracted the diamagnetic constants $\alpha$ of the transitions using the formula: $(E_n^{\sigma+}+E_n^{\sigma-})/2=E_{n,0}+\alpha_nB^2$ and the resulted values are indicated in Figure~\ref{fig:fig_energies}. Here, $E_n^{\sigma\pm}$ are the energies of the $n$th excitonic states in $\sigma^{\pm}$ polarizations, $E_{n,0}$ is the energy of the $n$th excitonic state at zero magnetic field $B=0$, and $\alpha_n$ is the diamagnetic shift of $n$th excitonic state. 

The obtained diamagnetic constants for ML are consistent with previous experimental findings~\cite{Stier2018}. Given the similarity in values observed for BL, we interpret the states similarly to those of ML - as consecutive states within a single exciton series. However, for TL and QL, subsequent states do not exhibit a similarly large increase in $\alpha_n$ with increasing the order $n$ of the considered exciton excited state. As a result, we assign the first two lowest energy resonances of TL as the ground states of two distinct exciton series, denoted as 1$s^\alpha$ and 1$s^\beta$, next two features (2$s^\alpha$ and 3$s^\alpha$) as excited states within the first series, and last resonance (2$s^\beta$) as an excited state of the second series. Similarly, for QL, we assign the first four states as ground states of four exciton series or two pairs of nearly degenerate ones (1$s^\alpha$ and 1$s^\beta$), with the next two states representing degenerate excited states within these series (2$s^\alpha$ and 2$s^\beta$). However, to validate these interpretations and gain a deeper quantitative understanding, a comprehensive theoretical analysis of the exciton series in WSe$_2$ multilayers is essential, which we address in the subsequent section of this article.

\section{Theoretical description}

Excitons in 2H-stacked S-TMD multilayers are bound states of electron-hole excitations from the K$^\pm$ points of the first Brillouin zone of these crystals. The basic properties of these excitations can be derived following the $\mathbf{k\cdot p}$ theory of the S-TMD monolayers \cite{Arora2019, Slobodeniuk2019, Grzeszczyk2021, Gong2013}.

\begin{figure*}[t]
\centering
	\includegraphics[width=0.75\linewidth]{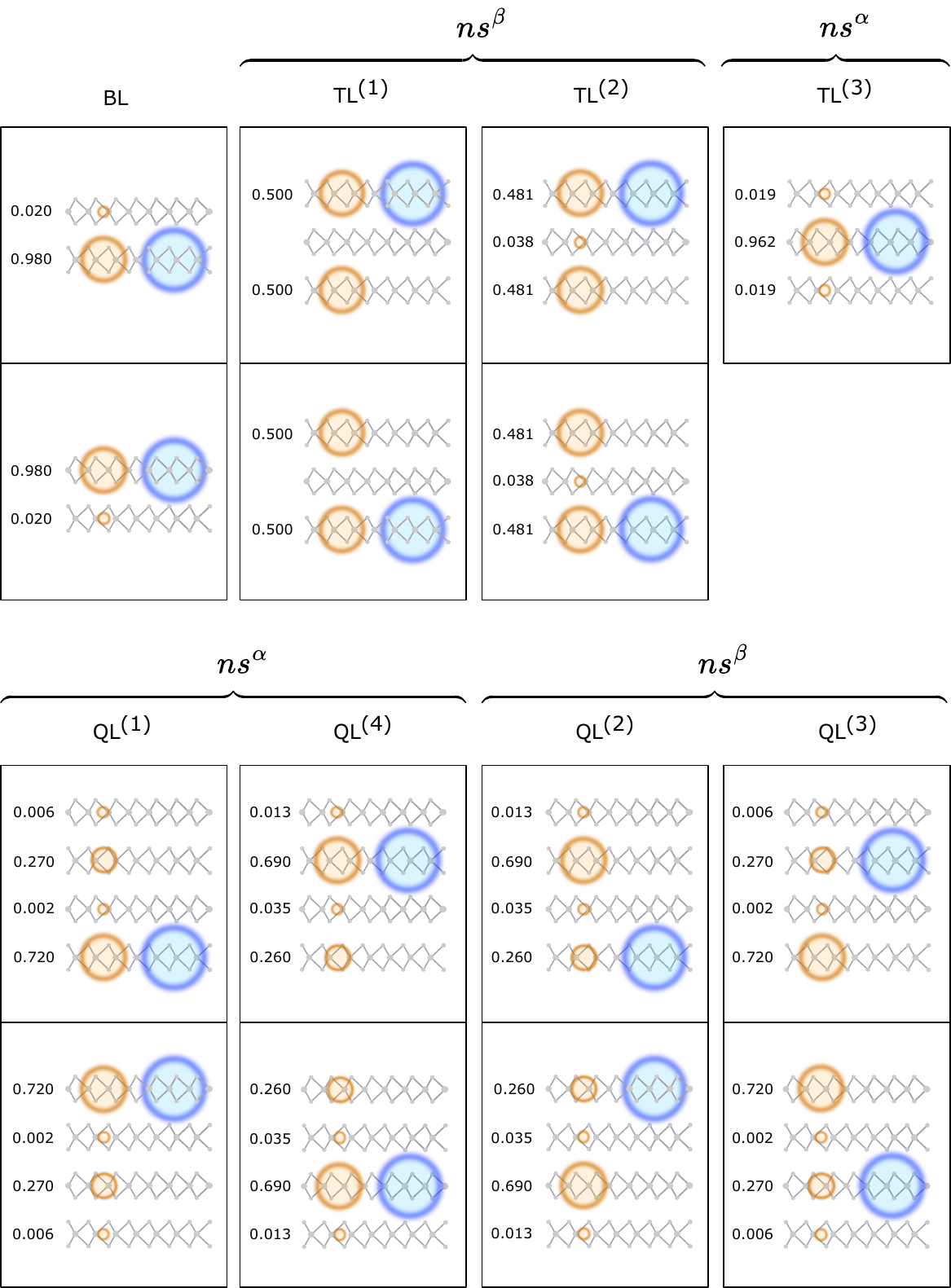}
	\caption{
                Schematical presentation of charge distribution of a hole (orange circles) and an electron (blue circles) in the 
                intralayer K-excitons in the bilayer (BL), 
                trilayer (TL$^{(1)}$, TL$^{(2)}$, TL$^{(3)}$), and quadlayer (QL$^{(1)}$, QL$^{(2)}$, QL$^{(3)}$, QL$^{(4)}$). 
                The numbers on the left side of each figure represent the amount of hole charge (in units $|e|$) 
                localized in each layer of the multilayer. Each panel shows a pair of images representing excitons with the same energies but mirrored electron and hole charge distributions, leading to opposite dipole moments, except for TL$^{(3)}$, which has a zero dipole moment.}
	\label{fig:full_picture_excitons}
\end{figure*}

According to the $\mathbf{k\cdot p}$ approach, the conduction bands (CB) of different layers of the multilayer are decoupled at their K$^\pm$ points.  
Hence the CB electrons of the multilayer can be considered in each layer separately. On the other hand, the valence bands (VB) of different layers of the multilayer are coupled at K$^\pm$ points. 
This results in the delocalization of VB state excitations between layers, leading to the delocalization of the hole charge of the corresponding hybridized VB in the direction perpendicular to the multilayer's planes. The charge distributions of the corresponding holes for bi-, tri-, and quadlayers are represented by orange circles in Fig.~\ref{fig:full_picture_excitons}. 
Note that the holes' charge distribution is not homogeneous within the layers. Namely, the layers containing prevailing amount of charge 
alternate with layers containing a minor amount of charge. Such behavior originates from the peculiarities of the interlayer coupling related to alternating orientation of layers and is valid for all multilayers of 2H stacking.     

The bound states of the holes and electrons correspond to so-called
K-excitons \cite{Slobodeniuk2019}. The K-excitons can be separated into two groups according to the location of their electrons with respect to the hole charge distribution in the multilayer -- ``intralayer'' and ``interlayer''. In the intralayer K-excitons, an electron is localized in one of the layers containing the prevailing amount of hole charge. Such excitons are characterized by strong oscillator strength and negative $g\approx -4$ factor. In the interlayer K-excitons, the electron remains in one of the layers containing a minor amount of hole charge. These excitons are characterized by a positive $g$-factor and weak oscillator strength. We focus on the optically dominant intralayer K-excitons, depicted in Fig.~\ref{fig:full_picture_excitons}.  

In the case of multilayers, the Coulomb potential between the delocalized hole and localized electron quasiparticles,
which form the intralayer K-exciton, deviates from the well-known Rytova-Keldysh potential \cite{Keldysh1979,Cudazzo2011,Slobodeniuk2023}
found in S-TMD monolayers. As a result, the excitons' binding energies are also modified compared to the monolayer case \cite{Kipczak_2023}. Therefore, to estimate the spectrum of the intralayer excitons in the multilayer structure one needs: i) to calculate the band structure of the hybridized VB states, and ii) to calculate the effective Coulomb potential between charges in the multilayer. The first part has been calculated analytically for simple cases of bi- and trilayer \cite{Slobodeniuk2019,Grzeszczyk2021} and numerically for $N=2,3,\dots 8$ \cite{Molas2017}. Here, we extend the previous studies and propose the general scheme of calculating the VB band structure in the multilayer, see SM for details~\cite{Supplementary_Materials}. To calculate the effective electron-hole Coulomb potential in the multilayer we extend the algorithm, developed for mono- and bilayer \cite{Kipczak_2023, Slobodeniuk2023}, to the arbitrary number of layers $N$.  The details of this calculation are also presented in SM~\cite{Supplementary_Materials}. 

Our method is based on the diagonalization of the electronic band Hamiltonian, introducing the new electron and hole
quasiparticles, and constructing the exciton states from these quasiparticles. Note that the alternative way of understanding the intravalley excitons in S-TMDs multilayers is presented in the literature 
\cite{Falko2019,Erkensten2023,Zhang2023,Lian2023,Feng2024}. In this approach, bare exciton states are introduced, where the electron and hole charges are fully localized in the specified layers of the multilayer system. The Hamiltonian of the crystal, expressed in the excitonic basis, is not diagonal. The off-diagonal matrix elements of this Hamiltonian represent the coupling between the bare exciton states, which arises from the interaction between the electronic bands of adjacent layers. The eigenstates of the excitonic Hamiltonian correspond to the actual excitonic states in the crystal, which are superpositions of the bare excitonic states (see detailed analysis in Ref.~\cite{Falko2019}). Although both approaches yield the same final results, our approach is less complex for intravalley excitons in multilayers with a 2H-stacking order. Therefore, we adopt this approach in our study. 

\subsection{Spectrum of excitons in S-TMD multilayers}

We consider the $N$-layer S-TMD crystal as a stack of monolayers arranged in parallel to $xy$ plane and 
separated equidistantly by distance $l$ in $z$ direction.
The multilayer is encapsulated in between hBN thick crystals with dielectric constant $\varepsilon=4.5$ \cite{Stier2018}.
Due to the crystal anisotropy, the S-TMD sample is polarized differently in the in- and the out-of-plane directions. 
The in-plane polarizability is defined by the 2D susceptibility, $\chi_{2D}=2\pi r_0$,
 of each monolayer constituting the multilayer. 
Here $r_0$ is an in-plane screening length of the monolayer. The dielectric constant $\epsilon_\perp$ defines the out-of-plane polarizability. 

In the considered $N$-layer crystal the motion of electrons and holes is restricted in the $xy$ plane. 
Therefore the properties of excitons, i.e., the bound states of electrons and holes, 
can be obtained from the effective two-dimensional Hamiltonian
\begin{align}
\label{eq:hamiltonian}
H^{(N)}=-\frac{\hbar^2}{2\mu}\nabla^2_\parallel+V^{(N)}_\text{eff}(\rho)
\end{align}
Here $\mu$ is the reduced effective mass of the exciton, $\nabla_\parallel=(\partial_\rho,\rho^{-1}\partial_\varphi)$ is the two-dimensional nabla operator, and $V_\text{eff}^{(N)}(\rho)$ is the effective Coulomb potential, between
an electron and a hole in the $N$-layer, as a function of the in-plane distance $\rho$ between the quasiparticles. 
Note that the hole's charge distribution in the $z$-direction is encoded in the form of the effective potential $V_\text{eff}^{(N)}(\rho)$.  

Solving the eigenvalue equation $H^{(N)}\psi^{(N)}(\rho,\varphi)=E^{(N)}\psi^{(N)}(\rho,\varphi)$ one can obtain the
binding energies $E^{(N)}_n$ and the wave-functions $\psi^{(N)}_n(\rho,\varphi)$ of the excitons with quantum number $n$ in the $N$-layer
crystal. We focus on $s$-excitonic states, whose wave function is angular-independent. Taking this fact into account and introducing the dimensionless parameter 
of the length $\xi=\rho\varepsilon\sqrt{\epsilon_\perp}/r_0$  we rewrite the eigenvalue equation in the dimensionless form 
\begin{equation}
\Big\{b^2\epsilon_\perp\frac{1}{\xi}\frac{d}{d\xi}\Big(\xi\frac{d}{d\xi}\Big)+
bv^{(N)}_\text{eff}(\xi)+\epsilon^{(N)}\Big\}\psi^{(N)}(\xi)=0.
\end{equation} 
Here $b=\hbar^2\varepsilon^2/\mu e^2r_0$, i.e., 
%$e$ is the electron's charge, and $\hbar$ is the Plank's constant. 
%This parameter is nothing but 
the ratio of the effective Bohr radius $a_0^*=\hbar^2\varepsilon/2\mu e^2$ and 
the reduced screening length $r_0^*=r_0/\varepsilon$. 
The dimensionless potential is defined as $v^{(N)}_\text{eff}(\xi)=-(r_0/e^2)V^{(N)}_\text{eff}(\xi r_0/\varepsilon\sqrt{\epsilon_\perp})$. 
Finally, the dimensionless energy parameter is determined as  $\epsilon^{(N)}=E^{(N)}/Ry^*$, 
where $Ry^*=\mu e^4/2\hbar^2\varepsilon^2=e^2/2a_0^*\varepsilon$ is the effective Rydberg energy in the system. 
We solve the dimensionless eigenvalue equation to analyze the properties of the excitons in WSe$_2$ mono-, bi-, tri-, 
and quadlayer crystals. 

For the case of ML the effective potential has the analytical form 
$v_\text{eff}^{(1)}(\xi)=\pi[\text{H}_0(\xi/\sqrt{\epsilon_\perp})-Y_0(\xi/\sqrt{\epsilon_\perp})]$,
where $\text{H}_0(x)$ and $Y_0(x)$ are the zero-order Struve and Neumann functions, respectively. 
Using the values $\mu=0.21m_0$ ($m_0$ is the electron's mass), the single particle band gap $E_\text{g}=1.873$~meV, 
$\epsilon_\perp=7.6$ \cite{Laturia2020}, $\varepsilon=4.5$, and $r_0=45$~\mbox{\AA} \cite{Stier2018}  
we reproduce the spectrum of the excitons in WSe$_2$ monolayer \cite{Molas2019}.  

\begin{figure*}[t]
	\centering
	\includegraphics[width=\linewidth]{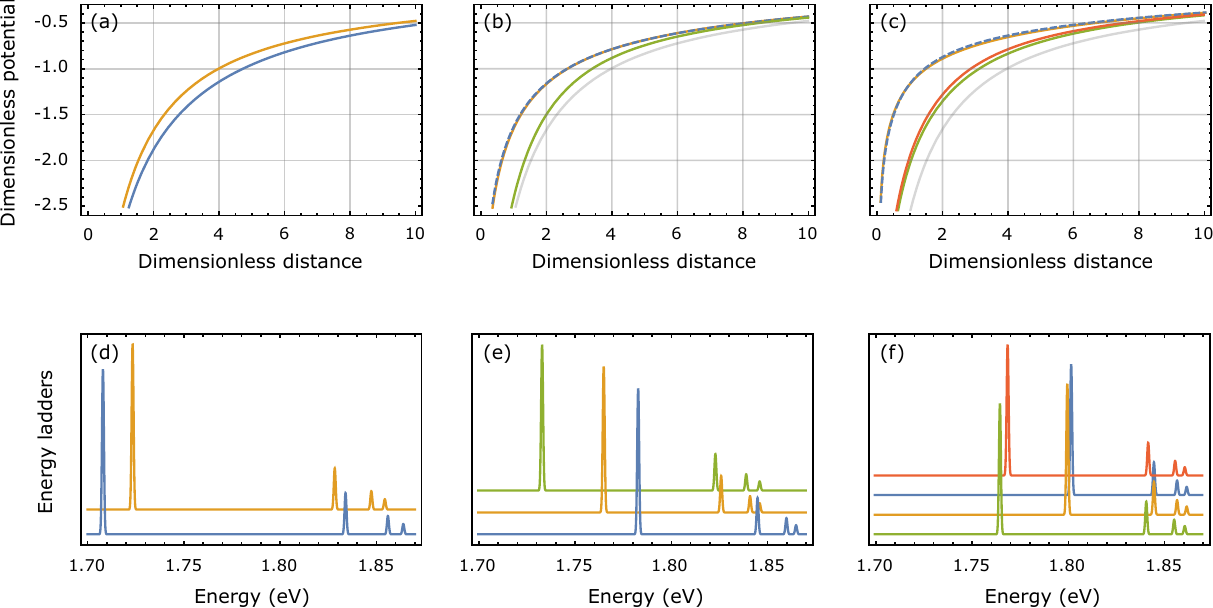}%
	\caption{ Effective potentials and excitonic spectra in the multilayers of S-TMDS. 
              (a) Effective dimensionless potentials for mono- ($-v_\text{eff}^{(1)}(\xi)$, blue curve) and 
	         bilayer ($-v_\text{eff}^{(2)}(\xi)$, yellow curve) 
	         as a function of dimensionless distance parameter $\xi=\rho\varepsilon\sqrt{\epsilon_\perp}/r_0$.
              (b) Effective dimensionless potentials for bi- ($-v_\text{eff}^{(2)}(\xi)$, gray reference curve) and 
	         trilayer ($-v_\text{eff}^{(3)}(\xi)$, yellow, dashed blue, and green curves). The potentials for TL$^{(1)}$ 
              (yellow curve) and TL$^{(2)}$ (dashed blue curve) have a similar, but not equal, shape.
              (c) Effective potentials for bi- ($-v_\text{eff}^{(2)}(\xi)$, gray reference curve) and 
	         quadlayer ($-v_\text{eff}^{(4)}(\xi)$ for $\text{QL}^{(1)}$ (green), $\text{QL}^{(2)}$ (yellow), 
              $\text{QL}^{(3)}$ (dashed blue), and $\text{QL}^{(4)}$ (red) curves). 
              Note that all the potentials are split into two groups with similar shapes.
              (d) Energy ladders of the excitons for mono- (blue curve) and 
	         bilayer (yellow curve), presented in the form of the absorption function. 
              The amplitude of the peaks represents the quantum number $n$ of the excitons -- 
              the smaller peak corresponds to the larger quantum number $n$.
              (e) Energy ladders of the excitons in trilayer, for antisymmetric (TL$^{(1)}$) excitons (blue curve), 
	         for symmetric excitons of the first (TL$^{(2)}$) and the second (TL$^{(3)}$) types (yellow and green curves, respectively).  
              (f) Energy ladders of the intralayer excitons in the quadlayer: $\text{QL}^{(1)}$ (green), $\text{QL}^{(2)}$ (yellow), 
              $\text{QL}^{(3)}$ (blue), and $\text{QL}^{(4)}$ (red) curves.}
	\label{fig:superpicture}
\end{figure*}

In the case of the BL the effective potential $v^{(2)}_\text{eff}(\xi)$ can't be written analytically.
We evaluate it numerically (see details in SM~\cite{Supplementary_Materials}) and present it in Fig.~\ref{fig:superpicture}~(a). 
As one can see the electron-hole potential in the bilayer is weaker than in the monolayer. 
This weakening is a manifestation of the redistribution of the hole wave function in the out-of-plane direction see left top panel in Fig.~\ref{fig:full_picture_excitons}. 
Note that the single-particle band gap (energy difference between the extrema of the VB and CB) 
for intralayer excitons in the bilayer is modified due to the interlayer coupling 
$E^\text{BL}_\text{g}=E_\text{g}+\Delta_v/2-\sqrt{\Delta_v^2/4+t^2}$ (see Ref.~\cite{Slobodeniuk2019} for details). 
Here $t\approx 67$~meV and $\Delta_v=456$~meV are the coupling parameter and spin-splitting in the VB of monolayers of WSe$_2$, respectively \cite{Gong2013}. 
Solving the eigenvalue problem for the bilayer and taking into account the modification of the band gap 
we derive the spectrum of the excitons in these crystals.  
It is presented in Fig.~\ref{fig:superpicture}~(d).
Note that the energy of the ground state exciton $n=1$ in the bilayer is larger than in the monolayer, while the situation is opposite for higher excited states $n>1$. Such behavior is a result of the competition between the change of the band gap and the reduction of the binding energies due to the weakening of the electron-hole potential in the bilayer. 
As mentioned, we utilized the monolayer's bandgap value $E_\text{g}$ to evaluate the spectrum of excitons in BL.
However, the parameter $E_\text{g}$ in the formula for the band gap of the bilayer ($E^\text{BL}_\text{g}$) can deviate from the ML. This deviation is due to the crystal field of the hBN substrate and adjacent layer. Therefore the relative positions of the ML and BL spectra, presented on 
Fig.~\ref{fig:superpicture}~(d), have an illustrative character. However, the difference of energies of the excited states ($E_n$, $n=2,3,\dots$) and ground state ($E_1$) excitons, i.e. where the band gap renormalization is compensated, can be used for comparison with the experimental results.  
Finally, the oscillator strength of the intralayer excitons in the bilayer is $I^\text{BL}=0.98 I^\text{ML}$, where $I^\text{ML}$ is the oscillator strength of the $A$-exciton line in 
the monolayer of WSe$_2$. 

The situation with the intralayer excitons in the TL is more complex. 
There are three different series of intralayer excitons in such crystals, which we marked as antisymmetric (TL$^{(1)}$), symmetric of the first (TL$^{(2)}$), and of the second (TL$^{(3)}$) type, see correspondent panels in Fig.~\ref{fig:full_picture_excitons}. The origin and the optical properties of the corresponding excitonic states are described in SM~\cite{Supplementary_Materials}.    
The set of the effective potentials, responsible for the formation of these three types of excitons is depicted in Fig.~\ref{fig:superpicture}~(b). 
Note that the potentials for TL$^{(1)}$ and TL$^{(2)}$ excitons almost coincide. Therefore the binding energies of these excitons are also close to each other. However, since the values of the single particle band gaps for the optical transitions that form these two types of excitons are different, their energy ladders do not overlap. Namely, the single-particle band gap for TL$^{(1)}$  exciton is not modified $E_\text{g}^{\text{TL}^{(1)}}=E_\text{g}$, while for 
symmetric excitons $E^{\text{TL}^{(2)}}_\text{g}=E^{\text{TL}^{(3)}}_\text{g}=E_\text{g}+\Delta_v/2-\sqrt{\Delta_v^2/4+2t^2}$ (see Ref.~\cite{Slobodeniuk2019} for details).  The numerically evaluated spectrum of the excitons in the trilayer is presented in Fig.~\ref{fig:superpicture}~(e). 
Note that the oscillator strengths of TL$^{(1)}$, TL$^{(2)}$, and TL$^{(3)}$ excitons are 
$I^{\text{TL}^{(2)}}\approx I^{\text{TL}^{(1)}}=I^\text{ML}$, $I^{\text{TL}^{(3)}}=0.962I^\text{ML}$, respectively. 

In the case of the QL four different optical transitions form the intralayer excitons, depicted as 
$|\Phi_{+,v}^{(-)}\rangle\rightarrow |\Psi_c^{(1)}\rangle$, $|\Phi_{+,v}^{(+)}\rangle\rightarrow |\Psi_c^{(1)}\rangle$, 
$|\Phi_{+,v}^{(-)}\rangle\rightarrow |\Psi_c^{(3)}\rangle$, and $|\Phi_{+,v}^{(+)}\rangle\rightarrow |\Psi_c^{(3)}\rangle$
in the current study.
Here the subscripts  in the $|\Phi_{+,v}^{(\pm)}\rangle$ mark the higher energy (``$+$'') VB states ($v$). 
The superscripts ``$\pm$'' distinguish between two different higher VB energy levels of the quadlayer. 
The states $|\Psi_c^{(m)}\rangle$, $m=1,3$ are the CB states of the electrons, localized in the $m$th layer.   
The additional information about the aforementioned states is presented in SM~\cite{Supplementary_Materials}.
Further, we enumerate these transitions, 
and the excitons related to them by indices 1,2,3, and 4 respectively. The oscillator strengths of these transitions are 
$I^\text{QL,1}=0.72I^\text{ML}$, $I^\text{QL,2}=0.26I^\text{ML}$, $I^\text{QL,3}=0.27I^\text{ML}$, and 
$I^\text{QL,4}=0.69I^\text{ML}$. 
These transitions appear due to the complex structure of the VB that emerges due to the interlayer coupling between the subsequent layers of the quadlayer. A detailed analysis of the interlayer interaction and its consequences are presented in 
SM~\cite{Supplementary_Materials}. The dimensionless potentials corresponding to four types of excitons are presented in Fig.~\ref{fig:superpicture}~(c).
These effective potentials are separated into two groups with similar shapes. As a result, the exciton spectrum in this system is also separated into two groups of spectral lines, 
see Fig.~\ref{fig:superpicture}~(f).  
The interlayer coupling between the VB of the QL produces four hybridized VB states -- two lower and two higher VB states. The intralayer excitons in the QL result from the optical transitions from the higher energy hybridized VB states to the CB ones. The single-particle band gaps for such exciton transitions are
$E_\text{g}^{\text{QL},\pm}=E_\text{g}+\Big(\Delta_v-\sqrt{\Delta_v^2+6t^2\pm 2\sqrt{5}t^2}\Big)/2$. 
Here sign ``$\pm$'' marks one of the VB states $|\Phi_{+,v}^{(\pm)}\rangle$ associated with these transitions.
Therefore, the groups of closest excitonic lines correspond to the potentials with similar shapes. These groups of excitons appear as doubled resonances in the experiment.

Let us now discuss the limitations of the presented model. In our study, we supposed that the energy position of the CB and VB in the multilayer is modified only due to the interlayer interaction between the layers. In this approximation, we calculate the energy positions of the CB and VB in the $N$-layer 
S-TMD crystal for an arbitrary number of layers $N$ (see the derivation in SM~\cite{Supplementary_Materials}). Using these values and calculating the binding energies of excitons in multilayers we estimate the excitonic spectra in the studied crystals. However, the energy positions of VB and CB are also affected by the crystal field of the hBN substrate and adjacent layers. As a result, the band gap in each layer of the crystal can deviate from the band gap of the monolayer. Additionally, the change of the spin-splitting of VB $\Delta_v$: i) causes the change of the hybridized VB positions, and ii) leads to the new distribution of the hole wave function in out-of-plane direction resulting in the modification of the effective Coulomb potential between an electron and hole. Finally, we should mention that the change in the positions of the CB and VB can be different for different layers depending on their distance to the hBN substrate. 

All shifts mentioned above can change the position of the excitonic lines in the spectra presented in Figs.~\ref{fig:superpicture}~(d),(e), and (f). Such an effect is the most pronounced in the trilayer, where the experiment demonstrates two excitonic 
lines instead of three, as predicted by the theory. In this case, one can suppose that two of the predicted lines are merged due to band gap modification, and/or interlayer coupling constant $t$ renormalization. We consider the possible influence of these effects in the ``Discussion'' section. 

To reduce the uncontrollable effect of the substrate we consider the difference between the binding energies $(E_n)$ of $n(=2,3,4)$th excited and of the ground $(E_1)$ intralayer exciton states $\Delta E_n=E_n-E_1$. These differences can be estimated from the experiment by taking the difference in the energies of the corresponding excitons. Such a procedure excludes partially the effects of the band gap renormalization due to the hBN substrate and allows us to compare the experimental results with theoretical estimations, presented in Table~\ref{tab:differencies_comp}. 

\begin{center}
\begin{table}[t]
 %\begin{center}
 \begin{tabular}{p{2.5cm} p{1.4cm}  p{1.4cm}  p{1.4cm} p{1.0cm}}
 \hline\hline \\ [-0.5ex]
 $\Delta E_n$\,[meV] & ML & BL  & TL$^{(m)}$ & QL$^{(m)}$  \\ [1.0ex]
 \hline \\
 $\Delta E_2$ (Theor.)  & 126  & 105 & 62$^{(1)}$   61$^{(2)}$   90$^{(3)}$  &  76$^{(1)}$    45$^{(2)}$   43$^{(3)}$    73$^{(4)}$ \\ [1.5ex]
 $\Delta E_2$ (Exp.)  & 128  & 100 & 88$^{\alpha}$ \quad 47$^{\beta}$ &  100$^{\alpha}$    54$^{\beta}$   44$^{\beta}$    87$^{\alpha}$ \\ [1.5ex]
 $\Delta E_3$ (Theor.)  & 148  & 124 & 77$^{(1)}$  76$^{(2)}$   106$^{(3)}$ &  91$^{(1)}$     57$^{(2)}$   55$^{(3)}$   87$^{(4)}$ \\ [1.5ex]
 $\Delta E_3$ (Exp.)  & 149  & 122 & 102$^{\alpha}$ &  - \\ [1.5ex]
 \hline
\end{tabular}
%\end{center}
\caption{\label{tab:differencies_comp}
Differences between the binding energies $(E_n)$ of $n(=2,3)$th excited and of the ground $(E_1)$ 
intralayer exciton states $\Delta E_n=E_n-E_1$ in the ML, BL, TL, and QL obtained theoretically (Theor.) and experimentally (Exp.). Parameter $m$ enumerates different exciton states in multilayers. }
\end{table}
\end{center}

\subsection{Zeeman and diamagnetic shifts of the intralayer excitons in multilayers of WSe2}

Let us discuss the Zeeman shifts of intralayer excitons considered above. Due to the spatial structure of these excitons, their $g$-factors should have the same sign as in the monolayer, i.e., the negative sign. 
This allows us to distinguish the intralayer K-excitons in the spectra from the interlayer ones (which $g$-factors are positive).  
However, the absolute values of $g$-factors of the excitons in the $N$ layers deviate from the monolayer case. 
This deviation appears from the complex structure of the hole excitations in multilayers. The theoretical analysis of exciton $g$-factors in multilayers can be found in Ref.~\cite{Slobodeniuk2019}. Note that the Zeeman shifts of excitons appear from the shifts of the bands in a magnetic field, see Ref.~\cite{Roth1959} for details. On the contrary, the diamagnetic shifts appear from the relative motion of an electron and a hole of the exciton in the presence of a magnetic field.  

Namely, for the constant out-of-plane magnetic field $\mathbf{B}=B\mathbf{e}_z$, the effective two-body Hamiltonian (\ref{eq:hamiltonian}) is modified into
\begin{equation}
H^{(N,B)}=H^{(N)}+\frac{\hbar|e|B}{2\mu c}L_z+\frac{e^2B^2}{8\mu c^2}\rho^2,
\end{equation}
where $c$ is the speed of light, and $L_z$ is the $z$-component of the angular momentum operator. 
The first term is the Hamiltonian of the two-body problem in the absence of the magnetic field. Its 
spectrum $E_n^{(N)}$ and the corresponding wave functions $\psi^{(N)}_n(\rho,\varphi)$ have been obtained previously.   
The two next terms provide the magnetic field contribution to the energies of the excitons in the $N$-layer. 
Since we focus on the $s$ excitonic states, which carry zero angular momentum, only the $B^2$ term gives the correction
$\Delta E_n^{(N)}$ to the spectrum  $E_n^{(N)}$. The leading term of this correction is  
\begin{equation}
\Delta E^{(N)}_n=2\pi\frac{e^2B^2}{8\mu c^2}\int_0^\infty d\rho \rho^3 |\psi_n^{(N)}(\rho)|^2=
\alpha_n^{(N)}B^2,
\end{equation} 
where we took into account that the wave functions of $s$-excitons are angular-independent $\psi_n(\rho,\varphi)=\psi_n(\rho)$. 
Rewriting the diamagnetic shifts $\alpha_n^{(N)}$ of $ns$-excitons in terms of dimensionless coordinates  
\begin{equation}
\alpha_n^{(N)}=\frac{e^2}{8\mu c^2} \frac{r_0^2}{\varepsilon^2\epsilon_\perp} 
\frac{\int_0^\infty d\xi \xi^3 |\psi^{(N)}_n(\xi)|^2}{\int_0^\infty d\xi \xi |\psi^{(N)}_n(\xi)|^2},
\end{equation}    
and taking the numerical value of the prefactor 
\begin{align}
\frac{e^2}{8\mu c^2}\frac{r_0^2}{\varepsilon^2\epsilon_\perp}
%=\frac{\mu_B^2}{Ry\epsilon_\perp}\frac{m_0}{\mu}\Big(\frac{r_0}{2a_0\varepsilon}\Big)^2
\approx 1.38\times 10^{-2}\mu\text{eV}/\text{T}^2,
\end{align}  
we evaluate numerically the diamagnetic shifts. They are presented in Table~\ref{tab:diam_shifts_comp}. 

\begin{center}
\begin{table}[t]
 \begin{tabular}{p{2.5cm} p{1.4cm}  p{1.4cm}  p{1.4cm} p{1.0cm}}
 \hline\hline \\ [-0.5ex]
 $\alpha_n^{(N)}$\,[$\mu$eV$\cdot$T$^{-2}$] & ML & BL  & TL$^{(m)}$ & QL$^{(m)}$  \\ [1.0ex]
 \hline \\
 $1s$ (Theor.)  & 0.28  & 0.32 & 0.60$^{(1)}$ 0.60$^{(2)}$ 0.36$^{(3)}$  &  0.45$^{(1)}$  0.90$^{(2)}$  0.90$^{(3)}$  0.50$^{(4)}$ \\ [1.5ex]
 $1s$ (Exp.)  & 1.1  & 0.3 & 0.5$^{\alpha}$ \quad  1.9$^{\beta}$  &  0.9$^{\alpha}$   2.2$^{\beta}$   2.8$^{\beta}$   1.0$^{\alpha}$ \\ [1.5ex]
 $2s$ (Theor.)  & 4.7   & 5.6  & 8.2$^{(1)}$ 8.2$^{(2)}$ 6.5$^{(3)}$ &  7.6$^{(1)}$  10.5$^{(2)}$ 10.6$^{(3)}$ 7.9$^{(4)}$ \\ [1.5ex]
 $2s$ (Exp.)  & 6.4   & 6.8  & 10$^{\alpha}$ \quad 14$^{\beta}$&   11$^{\alpha}$  5.5$^{\beta}$\\ [1.5ex]
 $3s$ (Theor.)  & 25    & 28   & 37$^{(1)}$ 37$^{(2)}$ 31$^{(3)}$    &  35$^{(1)}$ 44$^{(2)}$ 44$^{(3)}$ 36$^{(4)}$ \\  [1.5ex]
 $3s$ (Exp.)  & 23    & 19   & 34$^{\alpha}$  &  - \\  [1.5ex]
 \hline
\end{tabular}
\caption{\label{tab:diam_shifts_comp}
Diamagnetic shifts $\alpha_n^{(N)}$ of n$s$ $(n=1,2,3)$ intralayer exciton states  in the mono-(ML, $N=1$), bi-(BL, $N=2$), tri-(TL,$N=3$), and quadlayer (QL,$N=4$) obtained theoretically (Theor.) and experimentally (Exp.). Parameter $m$ enumerates different exciton states in multilayers. }
\end{table}
\end{center}

\section{Discussion}

\begin{figure}[t]
	\centering
	\includegraphics[width=\linewidth]{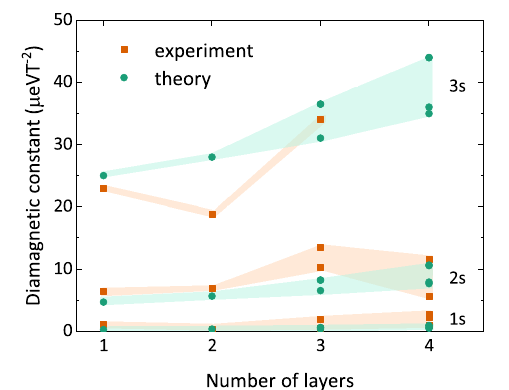}
	\caption{Comparison of experimentally obtained and theoretically estimated diamagnetic constants of observed exciton transitions as a function of the number of layers. The shaded regions serve as a visual guide.}
	\label{fig:exp_diamag}
\end{figure}

After determining the energies and diamagnetic constants of exciton resonances through experimental measurements and theoretical calculations, we compare the results to elucidate the origins of the observed emission lines in WSe$_2$ multilayers. Table~\ref{tab:differencies_comp} summarizes experimentally and theoretically derived differences in binding energies $(E_n)$ of $n$th excited states and the ground state $(E_1)$: $\Delta E_n = E_n - E_1$. The close agreement observed for ML and BL further validates our initial resonance assignments for these two structures. Notably, the theoretically predicted diamagnetic constants $\alpha_n$ (visually compared with experimentally obtained values in Fig.~\ref{fig:exp_diamag} and both listed in Table~\ref{tab:diam_shifts_comp}) exhibit close correspondence for ML and BL, resulting from the similar effective Coulomb potentials [see Fig.~\ref{fig:superpicture}~(a)].

In TL and QL WSe$_2$ multilayers, the excitonic structures become more intricate. Theoretical considerations suggest the presence of three and four distinct excitonic series in TL and QL, respectively. To reveal the origin of the observed features, we compare both their diamagnetic shifts and energy differences.

In TL, we identify five resonances: 1$s^\alpha$, 1$s^\beta$, 2$s^\alpha$, 3$s^\alpha$ and 2$s^\beta$, characterized by diamagnetic shifts of 0.5, 1.9, 10, 34, and 14 $\mu$eV$\cdot$T$^{-2}$, respectively. The first two relate to the ground states of two excitonic series. Among the three predicted series (TL$^{(1)}$, TL$^{(2)}$, and TL$^{(3)}$), TL$^{(1)}$ and TL$^{(2)}$ exhibit nearly degenerate diamagnetic constants and binding energy differences, whereas TL$^{(3)}$ displays significantly smaller diamagnetic shifts and larger binding energy differences. The problem of the number of excitonic states in the TL can be clarified using the recent studies of excitons in multilayer structures subjected to an out-of-plane electric field \cite{Zhang2023,Lian2023,Feng2024}. These studies indicate that the exciton with an energy $E=1.767$~eV, i.e. 1$s^\beta$ in our notation, possesses a dipole moment. 
Both TL$^{(1)}$, TL$^{(2)}$ K-excitons have identical out-of-plane dipole moments due to their similar electron-hole structure (see the corresponding panels in Fig.~\ref{fig:full_picture_excitons}). Consequently, their behavior in the presence of an out-of-plane electric field is indistinguishable. Based on the findings of Ref.~\cite{Zhang2023}, we conclude that both excitons should have the same energy. This can be achieved, for instance, by reducing the interlayer coupling constant $t$ and/or by renormalizing the band positions in the top and bottom layers of the trilayer (see discussion in the SI). Therefore, we consolidate the TL$^{(1)}$, and TL$^{(2)}$ lines with the same energy into a single TL$^{(1,2)}$ line. Note that the TL$^{(3)}$ exciton, which does not have an out-of-plane dipole moment, corresponds to $E=1.705$~eV line in Ref.~\cite{Zhang2023}. 

Based on diamagnetic shifts, we assign 1$s^\alpha$ to TL$^{(3)}$ and 1$s^\beta$ to TL$^{(1,2)}$. The energies of 2$s^\alpha$ and 3$s^\alpha$ are approximately 88 and 102 meV higher than the energy of 1$s^\alpha$, consistent with the predicted values for TL$^{(3)}$. The positioning of 2$s^\beta$, approximately 47 meV above 1$s^\beta$, suggests it is the first excited state of TL$^{(1,2)}$. The relative energies of two experimentally observed Rydberg series let us conclude that the n$s^\beta$ originates from merged 
antisymmetric TL$^{(1)}$ and symmetric of the first type TL$^{(2)}$ excitons [Fig.~\ref{fig:superpicture}~(e)].

For QL, we assign first pair of states marked as 1$s^\alpha$ to the theoretically predicted $\text{QL}^{(1)}$ and $\text{QL}^{(4)}$, and second pair 1$s^\beta$ to $\text{QL}^{(2)}$ and $\text{QL}^{(3)}$ based on absolute energies and diamagnetic constants (see Fig.~\ref{fig:superpicture}~(f) and Table~\ref{tab:diam_shifts_comp}). Similarly, we assign 2$s^\alpha$ to the degenerate excited states of $\text{QL}^{(1)}$ and $\text{QL}^{(4)}$, and 2$s^\beta$ to the degenerate excited states of $\text{QL}^{(2)}$ and $\text{QL}^{(3)}$ (see the corresponding charge distributions in Fig.~\ref{fig:full_picture_excitons}).

The presented theoretical model aligns with the experimental data at many points though we must admit the theory-experiment agreement is not always perfect, on the quantitative level in particular. We note, for example, that our model underestimates the measured diamagnetic constants in TL and QL samples as well as the difference $\Delta E_2$ in QL structure. Moreover, it does not capture well other experimental observations such as the decrease in the diamagnetic constant of the 3$s$ state in BL relative to ML or the observation of only two exciton series in the TL sample.

The presented theoretical modeling implies the use of several parameters. Our choice has been to fix their values, to the best of our knowledge, although the discrepancies regarding the actual values of these parameters are known in the literature. The theory-experiment agreement could be improved with an additional set of parameters (see the discussion in SM~\cite{Supplementary_Materials}), unfortunately not with a unique one. Therefore, we refrain from adding more parameters considering it too speculative, and remain satisfied with the actual understanding of the experimental data. The fair agreement with the experimental data obtained, nevertheless, allows us to give credence to the interpretation of the observed exciton complexes.

\section{Conclusions}

We have performed magneto-reflectance measurements on few-layer WSe$_2$ encapsulated in hBN structures. The new series of ground and excited excitonic resonances has been revealed in bi-, tri-, and quad-layers. The $g$-factors of the corresponding excitons have the same sign as in a monolayer. Therefore they belong to the class of the intralayer K-excitons. The measured diamagnetic shifts allow us to separate the ground excitonic states from the excited ones.

All the observed data can be explained if we suppose that the resonances result from i) the hybridizing valence and conduction band states between neighbor layers, and ii) modified Coulomb interaction between hybridized electronic states of the multilayer.
To verify this observation we derive the electronic properties of the multilayer within the $\mathbf{k\cdot p}$ approximation. 
The band structure of the valence and conduction bands has been evaluated as a function of the number of layers $N$. 
It was shown that the valance band of $N$-layer splits into $N$ subbands, among which some can be degenerate.
Due to this splitting $N$ different types of hole excitations exist in the monolayer, which gives rise to several different types of intralayer K-excitons. The energies of these excitons are a sum of two parts --- the single-particle band gap between new hybridized conduction and valence bands and the binding energy of the electron-hole pairs that form the excitons. The binding energies have been calculated separately using the developed theory of Coulomb interaction in multilayers.   

The developed theory i) qualitatively explains the structure of the excitonic lines in multilayers; ii) quantitatively explains the spectrum of excitons in mono- and bilayer; iii) quantitatively explain the diamagnetic shifts of the excitons in the multilayers.  

\section*{Acknowledgments}

The authors are grateful to D.~Vaclavkova and M.~Bartos for their collaboration at the initial stage of the work and thank B.~Pietka and M.~Kr\'ol for fruitful discussions. 
The work has been supported by the Czech Science Foundation (project GA\v{C}R 23-06369S). K.W. and T.T. acknowledge support from the JSPS KAKENHI (Grant Numbers 21H05233 and 23H02052) and World Premier International Research Center Initiative (WPI), MEXT, Japan. MP acknowledges the support from the Centera2 project, FENG.02.01-IP.05-T004/23, funded within the IRA program of the Foundation for Polish Science, co-financed by the EU FENG Programme.

P.K. and A.O.S. contributed equally to this work.

%\bibliography{main}


\newpage

\setcounter{figure}{0}
\setcounter{section}{0}

\newpage

\begin{widetext}
\begin{center}
	%%%%%%%%% ABSTRACT TITLE
	{\large{ {\bf Supplemental Material: Rydberg series of intralayer K-excitons in WSe$_2$ multilayers\\ }}}
	%%%%%%%%% ABSTRACT AUTHORS
	\vskip0.5\baselineskip{Piotr Kapuscinski,{$^{1,2}$} A. O. Slobodeniuk,{$^{3}$} Alex Delhomme,{$^{1,4}$} Cl\'{e}ment~Faugeras,{$^{1}$} Magdalena~Grzeszczyk,{$^{5}$} \linebreak[4] 
	Karol~Nogajewski,{$^{2}$} K. Watanabe,{$^{6}$} T. Taniguchi,{$^{7}$} and M. Potemski,{$^{1,8}$}}
	%%%%%%%%% AFFILIATION
	\vskip0.5\baselineskip{\em$^{1}$ Laboratoire National des Champs Magn\'etiques Intenses, LNCMI-EMFL, CNRS UPR3228,Univ. Grenoble Alpes, Univ. Toulouse, 
	 Univ. Toulouse 3, INSA-T, Grenoble and Toulouse, France \\
   $^{2}$ Institute of Experimental Physics, Faculty of Physics,
University of Warsaw, ul. Pasteura 5, PL-02-093 Warsaw, Poland \\
$^{3}$Department of Condensed Matter Physics, Faculty of Mathematics and Physics, Charles University, Ke Karlovu 5, CZ-121 16 Prague, Czech Republic \\
$^{4}$ Walter Schottky Institut and TUM School of Natural Sciences, Technische Universit{\"a}t M{\"u}nchen, Am Coulombwall 4, 85748 Garching, Germany \\
$^{5}$ Institute for Functional Intelligent Materials, National University of Singapore, Singapore 117544, Singapore \\
$^{6}$ Research Center for Electronic and Optical Materials, National Institute for Materials Science, 1-1 Namiki, Tsukuba 305-0044, Japan \\
$^{7}$ Research Center for Materials Nanoarchitectonics, National Institute for Materials Science,  1-1 Namiki, Tsukuba 305-0044, Japan \\
$^{8}$ CEZAMAT, CENTERA Labs, Warsaw University of Technology, PL-02-822 Warsaw, Poland \\
}
\end{center}

This supplemental material provides: \ref{gfactors} the values of the $g$-factors of the excitonic resonances observed in WSe$_2$ multilayers; \ref{VB_spectrum} calculation of the singleparticle band structure of the multilayers; \ref{coulomb} calculation of the Coulomb potential in the multilayers; \ref{effective} calculation of the energies of the excitons in WSe$_2$ multilayers.

\end{widetext}

\section{g-factors of the excitonic resonances observed in WSe$_2$ multilayers}
\label{gfactors}

\begin{figure*}[t]
	\centering
	\includegraphics[width=17.6cm]{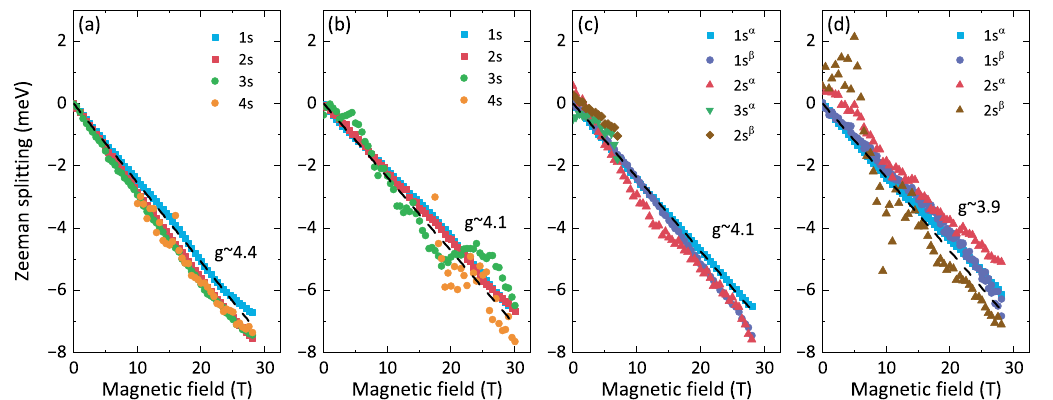}
	\caption{Zeeman splittings of the exciton resonances derived from the RC spectra of (a)~ML, (b)~BL, (c)~TL, and~(d) QL. }
	\label{fig:g-factors}
\end{figure*}

Figure~\ref{fig:g-factors} presets the values of Zeeman splitting of the excitonic transitions observed in magneto-reflectance experiments. G-factors of all the observeed transitions are negative, approximately $g \approx -4$, indicating that they originate from intralayer exciton transitions~\cite{aArora2017, aSlobodeniuk2019}.

\section{Single particle band structure of the conduction and valence bands of S-TMD multilayer}
\label{VB_spectrum}

We calculate the single-particle band structure using the symmetry arguments, previously considered in
Refs.~\cite{aSlobodeniuk2019, aMolas2017,aArora2019,aGrzeszczyk2021} for a case of bi- and tri-layers. 
We consider the general scheme and calculate the spectrum of the new conduction and valence bands in $N$-layer of 
transition metal dichalcogenides. 

Note that the considered quasi-two-dimensional crystals are characterized by 2H stacking order. It means that each upper layer
of the crystal is $180^\circ$ degrees rotated with respect to bottom one. This leads to the fact that the Brillouin zone 
of each upper layer is $180^\circ$ degrees rotated around the Brillouin zone of the previous bottom layer. Taking into account the 
hexagonal symmetry of the $N$-layer crystal we conclude that its Brillouin zone is also hexagonal and consists of $N$ Brillouin zones of the adjacent layers. We define the K$^+$ and K$^-$ points of the multilayer as the ones that coincide with the K$^+$ and K$^-$ points
of the bottom layer. Therefore the band structure at the K$^+$(K$^-$) point of the crystal is defined by the band structure  
at the K$^+$/K$^-$/K$^+$/K$^-\dots$ (K$^-$/K$^+$/K$^-$/K$^+\dots$) points of the monolayers in the $N$-layer. 
Further, we consider the electronic excitations at the K$^+$ point of the multilayer for brevity. 
The derivation for the K$^-$ point can be done analogously. 

The conduction band (CB) Bloch states of the multilayer in K$^+$ point are defined by the CB Bloch states of each layer 
$\{|\Psi_c^{(1)}\rangle|s\rangle,|\Psi_c^{(2)}\rangle|s\rangle,\dots |\Psi_c^{(N)}\rangle|s\rangle\}$. 
Here superscript $j=1,2,\dots N$ enumerates the number of the layer, and $s=\uparrow,\downarrow$ is spin index.
The symmetry of the states $|\Psi_c^{(j)}\rangle|s\rangle$ dictates that they are not mixed by the crystal field. 
It means that these states can be considered as the basis CB states of the multilayer. Therefore, the spectrum of the CB electron excitations is $N$ times degenerated. Namely the energy of the CB state $|\Psi_c^{(j)}\rangle|s\rangle$ of the $j$th layer is 
$E_c^{(j)}=E_c+(-1)^{j-1}\sigma_s\Delta_c/2$. Here $E_c$, $\Delta_c$ are the CB energy in the absence of the spin-orbit interaction, and CB spin-splitting of conduction bands in the monolayer, $\sigma_s=+1/-1$ for $s=\uparrow/\downarrow$ spin states.
Note that in the case of the additional anisotropy of the crystal in the out-of-plane direction, the CB states remain uncoupled. 
Hence, the degeneracy of the states in such a case lifts to $E_c^{(j)}=E_{c,j}+(-1)^{j-1}\sigma_s\Delta_{c,j}/2$, with the individual 
energies $E_{c,j}$, and the CB splitting $\Delta_{c,j}$ of $j$th layer. However, in the current study, we consider the degenerate situation for simplicity and discuss the effects of anisotropy at the end of the section.

To derive the fine structure of the valence band (VB)  of S-TMD $N$-layer in K$^+$ point we introduce the VB states $|\Psi_v^{(j)}\rangle|s\rangle$ in K$^+$ point of multilayer. In contrast to the CB states, the VB states are admixed by the 
electrostatic field of the crystal. The Hamiltonian, which describes the interaction between these states, 
written in the basis $\{|\Psi_v^{(1)}\rangle|s\rangle,|\Psi_v^{(2)}\rangle|s\rangle,\dots |\Psi_v^{(N)}\rangle|s\rangle\}$, is
$N\times N$ matrix
\begin{equation}
\mathcal{H}^{(N)}_{vs}=E_v\mathcal{I}+H^{(N)}_{vs}.
\end{equation}
Here $E_v$ is the energy of the VB state of the monolayer in the absence of spin-orbit interaction,  
$\mathcal{I}$ is a $N\times N$ unit matrix, and $H^{(N)}_{vs}$ is also a $N\times N$ unit matrix which
has a form 
 \begin{equation}
H^{(N)}_{vs}=\left[
\begin{array}{ccccc}
\sigma_s\frac{\Delta_v}{2} & t & 0  &\dots & 0 \\
t & -\sigma_s\frac{\Delta_v}{2} & t & \dots & 0\\
0 & t & \sigma_s\frac{\Delta_v}{2} & \dots  & 0\\ 
\vdots & \vdots & \vdots & \ddots &  \vdots  \\
\end{array}
\right].
\end{equation}   
Here $\Delta_v$ is the spin-splitting of the VB in the monolayer, and $t$ is the interlayer coupling parameter. 
Note that the signs of the diagonal matrix elements alternate. Hence in the case of an even number of layers $N=2M$, 
the last diagonal matrix element is $[H^{(N)}_{vs}]_{NN}=-\sigma_s\frac{\Delta_v}{2}$, while for an odd number of layers $N=2M+1$, $[H^{(N)}_{vs}]_{NN}=\sigma_s\frac{\Delta_v}{2}$. 
In the case of the out-of-plane anisotropy of the crystal, this Hamiltonian is modified. Namely, a) the scalar matrix 
$E_v\mathcal{I}$ should be replaced with the diagonal matrix $\text{diag}(E_{v,1}, E_{v,2}, \dots E_{v,N})$, where $E_{v,j}$ is 
the VB state energy of the $j$th layer;
b) the diagonal elements $[H^{(N)}_{vs}]_{jj}$ should be replaced with the diagonal matrix
$\text{diag}(\sigma_s\Delta_{v,1}/2, -\sigma_s\Delta_{v,2}/2, \dots )$, where $\Delta_{v,j}$ is 
the VB spin-splitting of the $j$th layer; and c) the off-diagonal matrix elements $[H^{(N)}_{vs}]_{j,j+1}=[H^{(N)}_{vs}]_{j+1,j}=t$ should be replaced with $t_j$. However, the evaluation of the spectrum and eigenstates of the general VB 
Hamiltonian, is impossible if all the matrix elements are unknown. Therefore, further, we consider the simple case as a reasonable approximation. We discuss the effects of anisotropy at the end of this section.      

The spectrum of the Hamiltonian can be found by solving the eigenvalue equation $J_N=\text{det}[H^{(N)}_{vs}-\lambda \mathcal{I}]=0$.
Note that the even $N=2M$ and odd $N=2M-1$ cases of this problem are different and the corresponding determinants 
$J_{2M}$ and $J_{2M-1}$ should be calculated separately. These determinants satisfy the following recurrence equations
\begin{align}
J_{2M}=&-(\sigma_s\Delta_v/2+\lambda)J_{2M-1}-t^2J_{2M-2},\\
J_{2M-1}=&(\sigma_s\Delta_v/2-\lambda)J_{2M-2}-t^2J_{2M-3}, 
\end{align}
with the conditions $J_0=1$, $J_1=\sigma_s\Delta_v/2-\lambda$. The structure of the recurrence relations 
dictates that $J_{2M-1}=-(\sigma_s\Delta_v/2-\lambda)R_{2M-1}$. Then we obtain 
\begin{align}
J_{2M}=&xR_{2M-1}-t^2J_{2M-2},\\
R_{2M+1}=&-J_{2M}-t^2R_{2M-1},
\end{align}
where $x\equiv\Delta_v^2/4-\lambda^2$. 
Note that the new recurrence relations don't depend on the spin sign $\sigma_s$. Therefore the spectrum of the energies $\lambda_j$, 
$j=1,2,\dots 2M$ for an even number of the layers  $N=2M$ doesn't depend on the spin, i.e. it is doubly degenerated by spin. 
This is in agreement with the time-reversal symmetry of the crystal. 
Moreover, $J_{2M}$ is the polynomial of $x$ and, hence,   
it is invariant under the transformation $\lambda\rightarrow -\lambda$. 
Hence the solution to the equation $J_{2M}=0$ has the structure $\lambda_l=\pm E_l$, 
$l=1,2,\dots M$. It reflects the VB structure of the even-numbered multilayer. 

Introducing the generation functions 
\begin{align}
J(z)=&\sum_{M=0}^\infty J_{2M} z^M,\\ 
R(z)=&\sum_{M=0}^\infty R_{2M+1}z^M,
\end{align}
and using the recurrence relations we obtain the following result
\begin{align}
J(z)[1+zt^2]-zxR(z)=1,\\ 
J(z)+R(z)[1+zt^2]=0.
\end{align}
The solution of these equations reads
\begin{align}
J(z)=\frac{1+zt^2}{[1+zt^2]^2+zx},\\ 
R(z)=-\frac{1}{[1+zt^2]^2+zx}.
\end{align}
Then, the determinant can be expressed as a contour integral around $\xi=zt^2=0$ point 
\begin{align}
J_{2M}=\frac{t^{2M}}{2\pi i}\oint \frac{d\xi}{\xi^{M+1}}\frac{1+\xi}{x^2+2\xi\Big(1+\frac{x}{2t^2}\Big)+1},\\
R_{2M+1}=\frac{t^{2M}}{2\pi i}\oint \frac{d\xi}{\xi^{M+1}}\frac{(-1)}{x^2+2\xi\Big(1+\frac{x}{2t^2}\Big)+1}.
\end{align}
Note that the parameter $x=\Delta_v^2/4-\lambda^2$ should be negative, due to the known phenomenon of the ``repelling" of 
energy levels of quantum systems in the presence of the interaction. Therefore it is convenient to use the parameterization $\cos\theta=1+x/2t^2$. 
Then  evaluating the integrals we get
\begin{align}
J_{2M}=(-1)^Mt^{2M}\frac{\cos\big(\frac{2M+1}{2}\theta\big)}{\cos(\theta/2)},\\
R_{2M+1}=(-1)^Mt^{2M}\frac{\sin\big([M+1]\theta\big)}{\sin\theta}.
\end{align}
The spectrum $\lambda_k$ of the VB of $2M$-layer can be obtained from the equation $J_{2M}=0$,  
\begin{align}
\lambda_k=\pm\sqrt{\frac{\Delta_v^2}{4}+4t^2\sin^2\Big(\frac{\pi}{2}\frac{2k+1}{2M+1}\Big)},
\end{align}
where $k=0,1,\dots, M-1$.
The spectrum $\lambda_k$ of the VB of $2M+1$ layer can be obtained from the equation 
$J_{2M+1}=-(\sigma_s\Delta_v/2-\lambda)R_{2M+1}=0$. 
It gives a pair of spin-dependent energies $\lambda_s=\sigma_s\Delta_v/2$, 
and $2M$ states, doubly degenerated by spin  
\begin{align}
\lambda_k=\pm\sqrt{\frac{\Delta_v^2}{4}+4t^2\sin^2\Big(\frac{\pi}{2}\frac{k}{M+1}\Big)},
\end{align}
where $k=1,\dots, M$. 

We consider the positions of the VB and CB bands for particular cases of mono-, bi-, tri-, and quadrolayer in detail.
We discuss the main effect as well as the contributions from the anisotropy of the system in an out-of-plane direction.  

\subsection{Monolayer}
\begin{widetext}

\begin{figure}[t]
	\centering
	\includegraphics[width=0.75\linewidth]{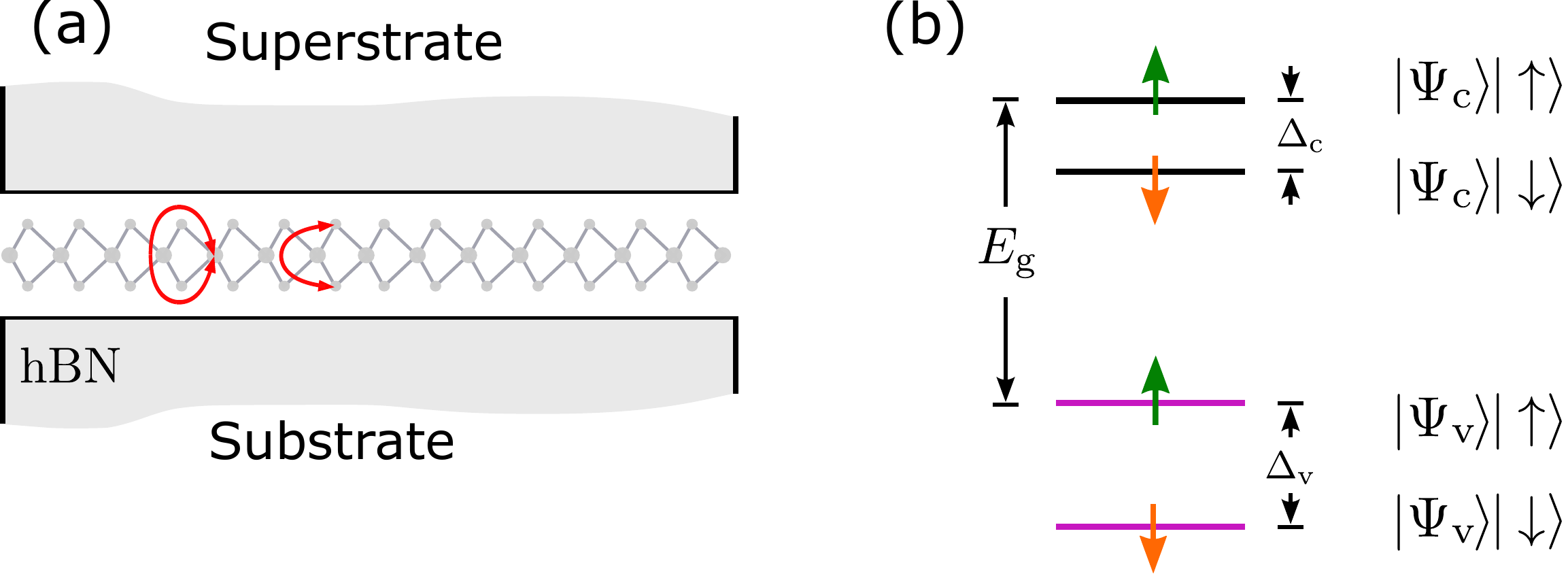}%
	\caption{ (a) The monolayer sample embedded in between two hBN substrates. The double-headed red arrows represent the in-plane mirror symmetry of the system. The position of the mirror plane coincides with the plane of the metal atoms of the crystal. (b) The band structure of WSe$_2$ monolayer at the K$^+$ point. The black/purple bar represents the spin-split conduction (CB) and valence (VB) bands. $\Delta_v$ and $\Delta_c$ represent the spin-splitting of the valence and conduction bands, respectively. $E_g$ is the difference of energies between spin-up conduction and valence band states, i.e. the lowest energy of optical transition in the considered monolayer. $|\Psi_v\rangle|s\rangle$ and $|\Psi_c\rangle|s\rangle$ states, with $s=\uparrow, \downarrow$, are the Bloch 
    VB and CB states in the K$^+$ point of the monolayer.} 
	\label{fig:monolayer_sample}
\end{figure}

\end{widetext}

We consider the CB and VB of the monolayer embedded in between the hBN medium, see Fig.~\ref{fig:monolayer_sample}.
The CB states are $|\Psi_c\rangle|\uparrow\rangle$, $|\Psi_c\rangle|\downarrow\rangle$ with energies 
$E_c+\Delta_c/2$ and $E_c-\Delta_c/2$, with $\Delta_c>0$. The VB states are $|\Psi_v\rangle|\uparrow\rangle$, $|\Psi_v\rangle|\downarrow\rangle$ with energies 
$E_v+\Delta_v/2$ and $E_v-\Delta_v/2$, with $\Delta_v>0$. The single-particle band gap in the system is 
$E_g=(E_c+\Delta_c/2)-(E_v+\Delta_v/2)$. Here $E_c$ and $E_v$ are positions of the CB and VB in the absence of 
the spin-orbit interaction. 

\subsection{Bilayer}
\begin{widetext}

\begin{figure}[t]
	\centering
	\includegraphics[width=0.85\linewidth]{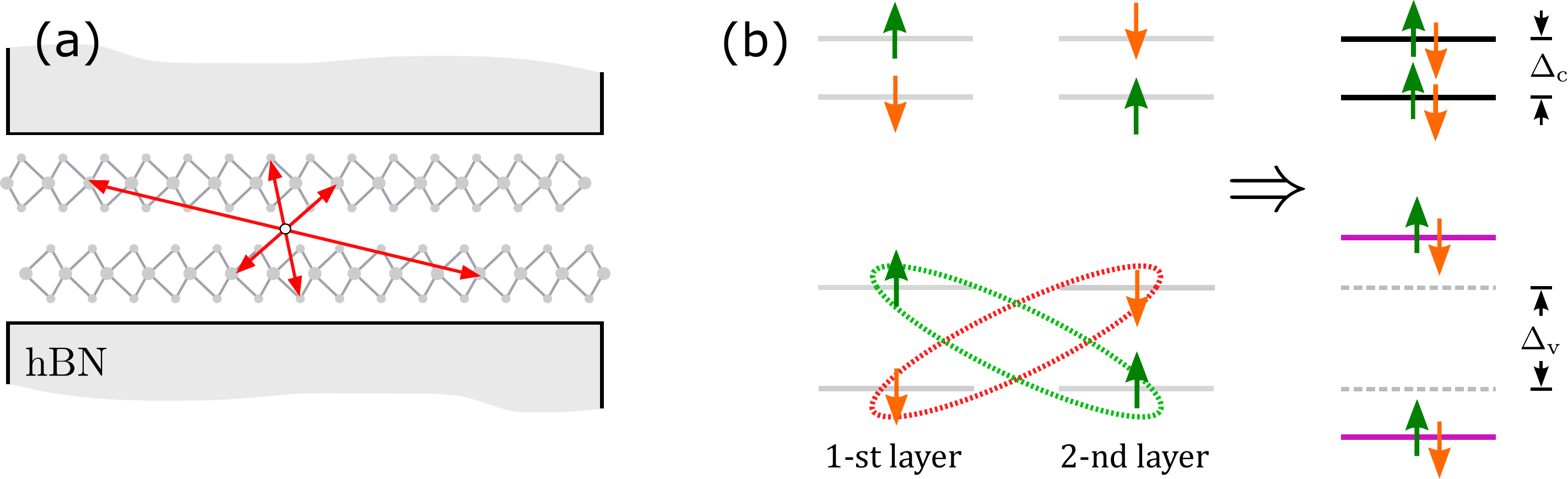}%
	\caption{(a) The bilayer sample embedded in between two hBN substrates. The double-headed red arrows represent the inverse symmetry of the system. The white circle represents the center of inversion. (b) The band structure of two separate layers with 2H stacking order (left-hand side). The positions of the VB and CB coincide due to the combination of the geometrical and time-reversal (TR) symmetries of the system. The band structure of WSe$_2$ bilayer at the K$^+$ point (right-hand side). The VB (purple bars) and CB (black bars) bands are doubly degenerated by spin due to the inverse and TR symmetries of the system. The splitting of the VB is larger than in the monolayer due to the repulsion of the energy levels. The repulsion appears from the coupling of the VB states with the same spin represented by green (for spin-up states) and orange (for spin-down states) ellipses. The strength of the coupling is defined by the parameter $t$. The splitting in the CB remains the same as in the monolayer due to specific symmetries of the CB Bloch states of the different layers in the bilayer. Due to geometry of the bilayer, the Bloch functions of different layers in the K$^+$ point are complex conjugated $\Psi_v^{(2)}(\mathbf{r})=\langle\mathbf{r}|\Psi_v^{(2)}\rangle=[\langle\mathbf{r}|\Psi_v^{(1)}\rangle]^*=
    [\Psi_v^{(1)}(\mathbf{r})]^*$, $\Psi_c^{(2)}(\mathbf{r})=\langle\mathbf{r}|\Psi_c^{(2)}\rangle=[\langle\mathbf{r}|\Psi_c^{(1)}\rangle]^*=[\Psi_c^{(1)}(\mathbf{r})]^*$.} 
	\label{fig:bilayer_sample}
\end{figure}
\end{widetext}

We consider the CB and VB of the bilayer embedded in between the hBN medium, see Fig.~\ref{fig:bilayer_sample}. 
According to the general theory the basis states of the CB and VB can be written as  
$\{|\Psi^{(1)}_c\rangle|s\rangle,|\Psi^{(2)}_c\rangle|s\rangle\}$
$\{|\Psi^{(1)}_v\rangle|s\rangle,|\Psi^{(2)}_v\rangle|s\rangle\}$, for spin-up $s=\uparrow$, and spin-down $s=\downarrow$ states 
separately. 

The CB Hamiltonian, written in the aforementioned basis
\begin{equation}
\mathcal{H}^{(2)}_{cs}=\left[
\begin{array}{cc}
E_c+\sigma_s\frac{\Delta_c}{2} & uk_+ \\
uk_- & E_c-\sigma_s\frac{\Delta_c}{2} \\
\end{array}
\right],
\end{equation}
where $E_g$ is the band gap of the bilayer and $k_\pm=k_x\pm ik_y$ are the in-plane momentum of conduction band quasiparticles. 
The $k_x,k_y$ dependent terms are small near the K$^+$ point, therefore we neglect them.  In this approximation, the CB states 
are defined by the diagonal matrix and hence do not interact with each other. As a result, the initial CB basis can be considered as a basis for the bilayer too. 

The VB Hamiltonian written on the aforementioned basis reads
\begin{equation}
\mathcal{H}^{(2)}_{vs}=\left[
\begin{array}{cc}
E_v+\sigma_s\frac{\Delta_v}{2} & t  \\
t & E_v-\sigma_s\frac{\Delta_v}{2}  \\
\end{array}
\right],
\end{equation}
where $\sigma_s=+1(-1)$ for $s=\uparrow(\downarrow)$, and 
the parameter $t$ defines the coupling between valence bands
from different layers. 
The eigenenergies of the corresponding Hamiltonian are $E^\pm_v=E_v\pm\sqrt{\Delta_v^2/4+t^2}$. 
The higher-energy eigenstates are
\begin{align}
&|\Phi^+_{v\uparrow}\rangle=\Big[\cos\theta|\Psi^{(1)}_v\rangle + \sin\theta|\Psi^{(2)}_v\rangle\Big]|\uparrow\rangle,
\nonumber \\
&|\Phi^+_{v\downarrow}\rangle=\Big[\sin\theta|\Psi^{(1)}_v\rangle+
\cos\theta|\Psi^{(2)}_v\rangle\Big]|\downarrow\rangle.
\end{align}
Here $\cos(2\theta)=\Delta_v/\sqrt{\Delta_v^2+4t^2}$.
Note that transfers of the electron from $|\Phi^+_{v\uparrow}\rangle$ state to the $|\Psi^{(1)}_c\rangle|\uparrow\rangle$ state, 
and from $|\Phi^+_{v\downarrow}\rangle$ state to the $|\Psi^{(2)}_c\rangle|\downarrow\rangle$ state lead to formation of the 
intralayer excitons (localized mainly in the first and second layers, respectively). Due to the symmetry of the system both excitons 
have the same energy. Therefore, the intralayer excitons in the bilayer are characterized by {\it single} exciton line. 
The degeneracy of intralayer exciton states appears from the similarity of the bottom and top layers of the bilayer from the electrostatic point of view. 

Let us discuss the difference between the bilayer and monolayer case. The electrostatic field of both hBN substrates acts on the monolayer. At the same time, each layer of the bilayer ``feels'' different electrostatic fields of the hBN substrates as well as the electrostatic field of the adjacent layer. Therefore the full CB and VB Hamiltonians of the bilayer get the corrections  
\begin{equation}
\delta\mathcal{H}^{(2)}_{cs}=\left[
\begin{array}{cc}
\delta E_c+\sigma_s\frac{\delta\Delta_c}{2} & 0 \\
0 & \delta E_c-\sigma_s\frac{\delta\Delta_c}{2} \\
\end{array}
\right],
\end{equation}
and 
\begin{equation}
\delta\mathcal{H}^{(2)}_{vs}=\left[
\begin{array}{cc}
\delta E_v+\sigma_s\frac{\delta \Delta_v}{2} & 0  \\
0 & \delta E_v-\sigma_s\frac{\delta \Delta_v}{2}  \\
\end{array}
\right],
\end{equation}
respectively. Note that the structure of the corrections coincides with the structure of the previous Hamiltonians. 
Therefore, all the previous conclusions remain the same. All the numerical results can be obtained from the previous ones by simple substitution $E_c\rightarrow E_c+\delta E_c$, $E_v\rightarrow E_v+\delta E_v$, $\Delta_c\rightarrow \Delta_c+\delta \Delta_c$, and 
$\Delta_v\rightarrow \Delta_v+\delta \Delta_v$. 

\subsection{Trilayer}

\begin{widetext}

\begin{figure}[t]
	\centering
	\includegraphics[width=\linewidth]{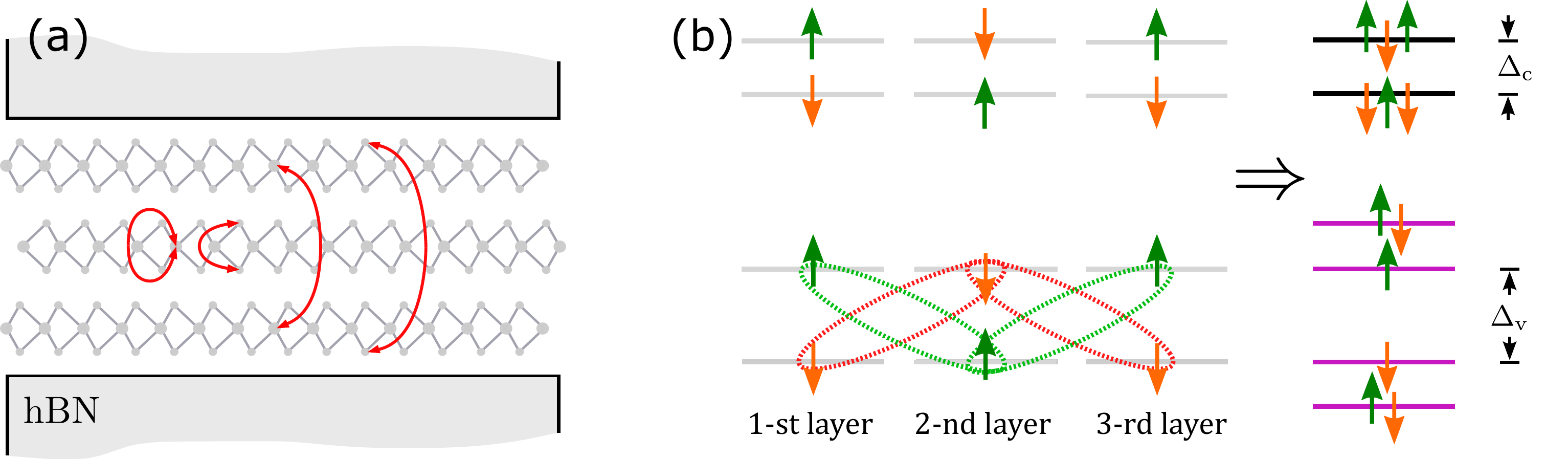}%
	\caption{(a) The trilayer sample embedded in between two hBN substrates. The double-headed red arrows represent the in-plane mirror symmetry of the system. The position of the mirror plane coincides with the plane of metal atoms of the middle layer of the WSe$_2$. (b) The left-hand side of the picture represents the band structure of non-interacting layers with 2H stacking order. The band structure of WSe$_2$ trilayer at the K$^+$ point (right-hand side). The symmetry of the system dictates the existence of the even and odd states under the mirror symmetry transformation. The even VB states (purple bars)  and CB (black bars) are doubly degenerated by spin, while the odd VB and CB states are spin polarized. The coupling of the VB states with the same spin is represented by green (for spin-up states) and orange (for spin-down states) ellipses. The strength of the coupling is defined by the parameter $t$. The splitting in the CB remains the same as in the monolayer due to specific symmetries of the CB Bloch states of the different layers in the trilayer. Due to the geometry of the crystal, the Bloch functions of different layers in the K$^+$ point are complex and satisfy the following relations $\Psi_v^{(2)}(\mathbf{r})=[\Psi_v^{(1)}(\mathbf{r})]^*=[\Psi_v^{(3)}(\mathbf{r})]^*$, $\Psi_c^{(2)}(\mathbf{r})=[\Psi_c^{(1)}(\mathbf{r})]^*=[\Psi_c^{(3)}(\mathbf{r})]^*$.} 
	\label{fig:trilayer_sample}
\end{figure}
\end{widetext}

Here we discuss the case of the trilayer sample depicted in Fig.~\ref{fig:trilayer_sample}.
The Hamiltonian for conduction bands, written on the basis
$\{|\Psi^{(1)}_c\rangle|s\rangle,|\Psi^{(2)}_c\rangle|s\rangle,|\Psi^{(3)}_c\rangle|s\rangle\}$, reads
\begin{equation}
\mathcal{H}^{(3)}_{cs}=\left[
\begin{array}{ccc}
E_c+\sigma_s\frac{\Delta_c}{2} & 0 & 0 \\
0 & E_c-\sigma_s\frac{\Delta_c}{2} & 0 \\
0& 0&  E_c+\sigma_s\frac{\Delta_c}{2}\\
\end{array}
\right].
\end{equation}

The Hamiltonian for VB states, written in the basis
$\{|\Psi^{(1)}_v\rangle|s\rangle,|\Psi^{(2)}_v\rangle|s\rangle,
|\Psi^{(3)}_v\rangle|s\rangle\}$, is
\begin{equation}
\mathcal{H}^{(3)}_{vs}=\left[
\begin{array}{ccc}
E_v+\sigma_s\frac{\Delta_v}{2} & t & 0  \\
t & E_v-\sigma_s\frac{\Delta_v}{2} & t  \\
0 & t & E_v+\sigma_s\frac{\Delta_v}{2}  \\
\end{array}
\right].
\end{equation}

The corrections to these Hamiltonians are more complicated than in the bilayer case. Indeed the contributions of the crystal field of the hBN sub- and superstrates, as well as the contributions from the crystal field of the adjacent layers are different for top/bottom 
and middle layers. Namely, the symmetry of the system dictates that the corrections to the energies $E_c$, $E_v$, and splitting 
$\Delta_c$, $\Delta_v$ are equal for the top and bottom layers, and deviate from the same type of corrections for the middle layer.
The interlayer coupling $t$ for the trilayer should also deviate from its bilayer value, i.e. $t\rightarrow t+\delta t$. This correction has the same value for interlayer coupling between the bottom and middle layers as well as for the interlayer coupling between the middle and top layers. This conclusion also comes from the symmetry of the system.
Taking this observation into account, we obtain the correction terms to the CB and VB Hamiltonians of the system 
\begin{widetext}
\begin{align}
\delta\mathcal{H}^{(3)}_{cs}=&\left[
\begin{array}{ccc}
\delta E_{c,1}+\sigma_s\frac{\delta\Delta_{c,1}}{2} & 0 & 0 \\
0 & \delta E_{c,2}-\sigma_s\frac{\delta\Delta_{c,2}}{2} & 0 \\
0& 0 &  \delta E_{c,1}+\sigma_s\frac{\delta\Delta_{c,1}}{2}\\
\end{array}
\right],
\end{align}
\begin{align}
\delta \mathcal{H}^{(3)}_{vs}=&\left[
\begin{array}{ccc}
\delta E_{v,1}+\sigma_s\frac{\Delta_{v,1}}{2} & \delta t & 0  \\
\delta t & \delta E_{v,2}-\sigma_s\frac{\Delta_{v,2}}{2} & \delta t  \\
0 & \delta t & \delta E_{v,1}+\sigma_s\frac{\delta \Delta_{v,1}}{2}  \\
\end{array}
\right].
\end{align}
\end{widetext}
The system possesses a mirror symmetry. Hence, the eigenstates of the VB and CB Hamiltonians can be classified 
by the irreducible representations of this symmetry. Namely, the states should be either odd or even under mirror transformation.
Therefore, it is convenient to introduce the basis states which possesses this symmetry 
\begin{align}
|\Upsilon_{ns}^{(\text{as})}\rangle=&\frac{1}{\sqrt{2}}\big[|\Psi^{(1)}_n\rangle - |\Psi^{(3)}_n\rangle\big]|s\rangle, \\
|\Upsilon_{ns}^{(\text{s,I})}\rangle=&\frac{1}{\sqrt{2}}\big[|\Psi^{(1)}_n\rangle + |\Psi^{(3)}_n\rangle\big]|s\rangle, \\
|\Upsilon_{ns}^{(\text{s,II})}\rangle=&|\Psi^{(2)}_n\rangle|s\rangle,
\end{align}
where $n=c,v$. Rewriting the full Hamiltonians on the new basis we obtain the CB and VB Hamiltonians in the following form 
\begin{widetext}
\begin{equation}
\mathcal{H}^{(3)}_{cs}+\delta \mathcal{H}^{(3)}_{cs}=\left[
\begin{array}{ccc}
E_c+\delta E_{c,1}+\sigma_s\frac{\Delta_c+\Delta_{c,1}}{2} & 0 & 0 \\
0 & E_c+\delta E_{c,1}+\sigma_s\frac{\Delta_c+\Delta_{c,1}}{2} & 0 \\
0 & 0&  E_c+\delta E_{c,2}-\sigma_s\frac{\Delta_c+\Delta_{c,2}}{2}\\
\end{array}
\right],
\end{equation}
\begin{equation}
\mathcal{H}^{(3)}_{vs}+\delta \mathcal{H}^{(3)}_{vs}=\left[
\begin{array}{ccc}
E_v+\delta E_{v,1}+\sigma_s\frac{\Delta_v+\Delta_{v,1}}{2} & 0 & 0 \\
0 & E_v+\delta E_{v,1}+\sigma_s\frac{\Delta_v+\Delta_{v,1}}{2} & \sqrt{2}(t+\delta t) \\
0 & \sqrt{2}(t+\delta t) &  E_v+\delta E_{v,2}-\sigma_s\frac{\Delta_v+\Delta_{v,2}}{2}\\
\end{array}
\right].
\end{equation}
\end{widetext}
One can see that the antisymmetric states are decoupled from the symmetric ones. The optical transitions 
are allowed only between the VB and CB states of the same parity. Therefore, one can tell about the antisymmetric 
and symmetric transitions. The Coulomb-bound electron-hole pairs that appeared due to these transitions, 
can be called also antisymmetrical and symmetrical excitons. 

Note that the energies of the transitions are independent due to the corrections $\delta E_{n,1}$, $\delta E_{n,2}$, 
$\delta \Delta_{n,1}$, $\delta \Delta_{n,2}$, with $n=c,v$ and $\delta t$. Therefore the relative positions of the corresponding
exciton lines can be estimated if only all the aforementioned corrections are known.  

\subsection{Quadrolayer}

Here we consider the quadrolayer sample embedded in between two hBN media, see Fig.~\ref{fig:quadrolayer_sample}.  
The leading parts of Hamiltonians for CB and VB, written on the basis
$\{|\Psi^{(1)}_n\rangle|s\rangle,|\Psi^{(2)}_n\rangle|s\rangle,|\Psi^{(3)}_n\rangle|s\rangle,|\Psi^{(4)}_n\rangle|s\rangle\}$, 
with $n=c,v$, read
\begin{widetext}
\begin{equation}
\mathcal{H}^{(4)}_{cs}=\left[
\begin{array}{cccc}
E_c+\sigma_s\frac{\Delta_c}{2} & 0 & 0 & 0 \\
0 & E_c-\sigma_s\frac{\Delta_c}{2} & 0 & 0 \\
0 & 0 &  E_c+\sigma_s\frac{\Delta_c}{2} & 0\\
0 & 0 &  0 &  E_c-\sigma_s\frac{\Delta_c}{2} 
\end{array}
\right].
\end{equation}
\begin{equation}
\mathcal{H}^{(4)}_{vs}=\left[
\begin{array}{cccc}
E_v+\sigma_s\frac{\Delta_v}{2} & t & 0 & 0  \\
t & E_v-\sigma_s\frac{\Delta_v}{2} & t & 0 \\
0 & t & E_v+\sigma_s\frac{\Delta_v}{2} & t  \\
0& 0 & t & E_v-\sigma_s\frac{\Delta_v}{2} 
\end{array}
\right].
\end{equation}

\begin{figure}[t]
	\centering
	\includegraphics[width=\linewidth]{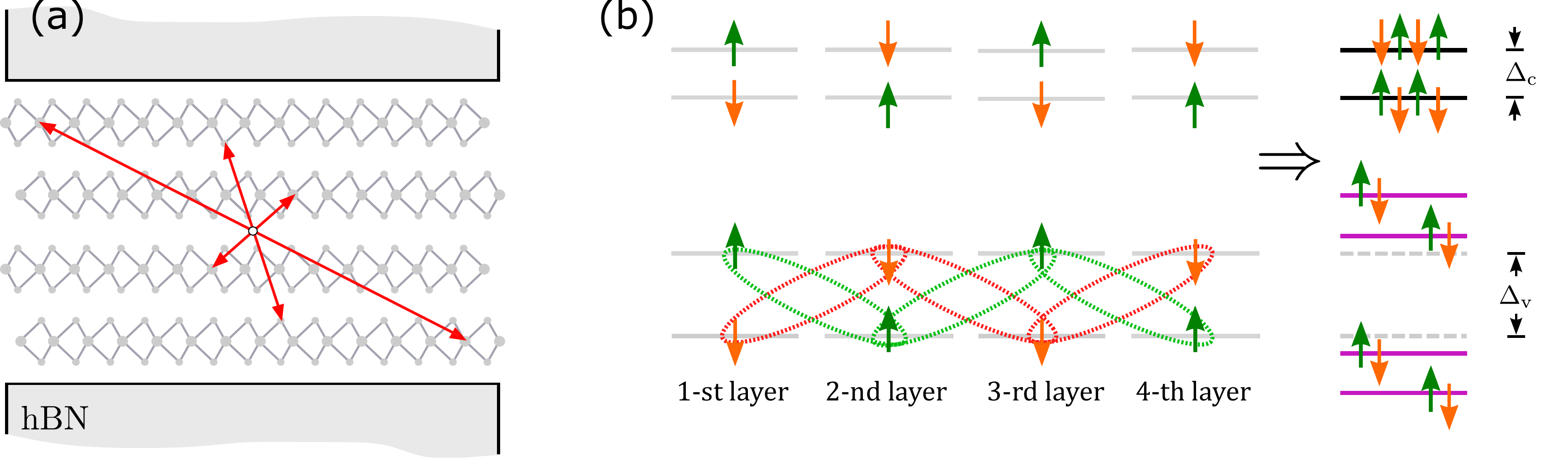}%
	\caption{(a) The quadrolayer sample embedded in between two hBN substrates. The double-headed red arrows represent the inverse symmetry of the system. The white circle represents the center of inversion. (b) The left-hand side of the picture represents the band structure of non-interacting layers with 2H stacking order. The band structure of WSe$_2$ quadrolayer at the K$^+$ point (right-hand side). The symmetries of the system dictate the existence of double degenerated by spin states. The coupling of the VB states with the same spin is represented by green (for spin-up states) and orange (for spin-down states) ellipses. The strength of the coupling is defined by the parameter $t$. The splitting in the CB remains the same as in the monolayer due to specific symmetries of the CB Bloch states of the different layers in the trilayer. Due to the geometry of the crystal, the Bloch functions of different layers in the K$^+$ point are complex and satisfy the following relations $\Psi_v^{(2)}(\mathbf{r})=[\Psi_v^{(1)}(\mathbf{r})]^*=[\Psi_v^{(3)}(\mathbf{r})]^*=\Psi_v^{(4)}(\mathbf{r})$, $\Psi_c^{(2)}(\mathbf{r})=[\Psi_c^{(1)}(\mathbf{r})]^*=[\Psi_c^{(3)}(\mathbf{r})]^*=\Psi_c^{(4)}(\mathbf{r})$.} 
	\label{fig:quadrolayer_sample}
\end{figure}
\end{widetext}

The crystal field corrections to these Hamiltonians, which take into account the symmetry of the quadrolayer system reads
\begin{widetext}
\begin{equation}
\delta \mathcal{H}^{(4)}_{cs}=\left[
\begin{array}{cccc}
\delta E_{c,1}+\sigma_s\frac{\Delta_{c,1}}{2} & 0 & 0 & 0 \\
0 & \delta E_{c,2}-\sigma_s\frac{\Delta_{c,2}}{2} & 0 & 0 \\
0 & 0 &  \delta E_{c,2}+\sigma_s\frac{\Delta_{c,2}}{2} & 0\\
0 & 0 &  0 &  \delta E_{c,1}-\sigma_s\frac{\Delta_{c,1}}{2} 
\end{array}
\right].
\end{equation}
\begin{equation}
\delta \mathcal{H}^{(4)}_{vs}=\left[
\begin{array}{cccc}
\delta E_{v,1}+\sigma_s\frac{\Delta_{v,1}}{2} & \delta t_1 & 0 & 0  \\
\delta t_1 & \delta E_{v,2}-\sigma_s\frac{\delta \Delta_{v,2}}{2} & \delta t_2 & 0 \\
0 & \delta t_2 & \delta E_{v,2}+\sigma_s\frac{\delta \Delta_{v,2}}{2} & \delta t_1  \\
0 & 0 & \delta t_1 & \delta E_{v,1}-\sigma_s\frac{\delta \Delta_{v,1}}{2} 
\end{array}
\right].
\end{equation}
\end{widetext}
Contrary to the trilayer case we can't introduce the new convenient basis states that diagonalize the full Hamiltonians. 
The corresponding spectrum and eigenstates should be obtained from the general scheme. However, the inverse symmetry of the
system dictates that as a minimum each band should be doubly degenerated by spin. The symmetry analysis of the CB states 
manifests that each CB band is 4 times degenerated, see Fig.~\ref{fig:quadrolayer_sample}~(b). 
This conclusion can be easily observed from a diagonal form of the full CB Hamiltonian $\mathcal{H}^{(4)}_{cs}+\delta \mathcal{H}^{(4)}_{cs}$ in the system.  For the case of VB, the inverse symmetry dictates that each band is as minimum doubly degenerated by spin. This result can be obtained by analyzing the matrices for the spin-up  $\mathcal{H}^{(4)}_{v\uparrow}+\delta \mathcal{H}^{(4)}_{v\uparrow}$ and spin-down $\mathcal{H}^{(4)}_{v\downarrow}+\delta \mathcal{H}^{(4)}_{v\downarrow}$ in the K$^+$
point of the quadrolayer. Namely, spin-up and spin-down Hamiltonians have the same matrix form being written on the bases 
$\{|\Psi^{(1)}_v\rangle|\uparrow\rangle,|\Psi^{(2)}_v\rangle|\uparrow\rangle,|\Psi^{(3)}_v\rangle|\uparrow\rangle,
|\Psi^{(4)}_v\rangle|\uparrow\rangle\}$ and 
$\{|\Psi^{(4)}_v\rangle|\downarrow\rangle,|\Psi^{(3)}_v\rangle|\downarrow\rangle,|\Psi^{(2)}_v\rangle|\downarrow\rangle,
|\Psi^{(1)}_v\rangle|\downarrow\rangle\}$, respectively. Therefore both Hamiltonians have the same spectrum of energies, 
and hence each state of VB, originating from both Hamiltonias, is doubly degenerated by spin. 
The relative positions of VB of the quadrolayer can be qualitatively understood by the following procedure. 
It is enough to calculate the positions of the bands for spin-up states since spin-down states will have 
the same positions.  
We rewrite the full Hamiltonian  $\mathcal{H}^{(4)}_{v\uparrow}+\delta \mathcal{H}^{(4)}_{v\uparrow}$ on the basis 
$\{|\Psi^{(1)}_v\rangle|\uparrow\rangle,|\Psi^{(3)}_v\rangle|\uparrow\rangle,|\Psi^{(4)}_v\rangle|\uparrow\rangle,
|\Psi^{(2)}_v\rangle|\uparrow\rangle\}$, to separate the higher and lower VB existing in each layer 
of the quadrolayer without the interlayer coupling. The Hamiltonian then reads
\begin{equation}
\mathcal{H}^{(4)}_{v\uparrow}+\delta \mathcal{H}^{(4)}_{v\uparrow}=\Big(E_v+\frac{\delta E_{v,1}+\delta E_{v,2}}{2}\Big)\mathcal{I}+
\left[
\begin{array}{cc}
\Lambda_+ &  T  \\
T & \Lambda_-
\end{array}
\right],
\end{equation}
where we  $4\times 4$ unit matrix $\mathcal{I}$, and the following $2\times 2$ matrices 
\begin{widetext}
\begin{equation}
\Lambda_+=\left[
\begin{array}{cc}
\frac{\delta E_{v,1}-\delta E_{v,2}}{2}+\frac{\Delta _v+\Delta_{v,1}}{2} & 0   \\
0 & -\frac{\delta E_{v,1}-\delta E_{v,2}}{2}+\frac{\Delta _v+\Delta_{v,2}}{2}  
\end{array}
\right],
\end{equation}
\quad 
\begin{equation}
\Lambda_-=\left[
\begin{array}{cc}
 \frac{\delta E_{v,1}-\delta E_{v,2}}{2}-\frac{\Delta_v+\Delta_{v,1}}{2}  & 0  \\
0 & -\frac{\delta E_{v,1}-\delta E_{v,2}}{2}-\frac{\Delta_v+\Delta_{v,2}}{2}
\end{array}
\right],
\end{equation} 
\quad
\begin{equation}
T=\left[
\begin{array}{cc}
 0  &  t+\delta t_1  \\
 t+\delta t_1 & t+\delta t_2
\end{array}
\right].
\end{equation} 
\end{widetext}
The diagonal matrices $\Lambda_+$ and $\Lambda_-$ describe the energy positions of the upper and lower 
spin-up VB states without interlayer couplings, defined by the matrix $T$. Using the matrix $T$ as a small parameter 
we apply the Lowdin procedure to the Hamiltonian, partially diagonalize it, and obtain 
\begin{widetext}
\begin{equation}
\mathcal{H}^{(4)}_{v\uparrow}+\delta \mathcal{H}^{(4)}_{v\uparrow}\approx \Big(E_v+\frac{\delta E_{v,1}+\delta E_{v,2}}{2}\Big)\mathcal{I}+
\left[
\begin{array}{cc}
\Lambda_+ +T(\Lambda_+-\Lambda_-)^{-1}T & 0  \\
0 & \Lambda_- -T(\Lambda_+-\Lambda_-)^{-1}T
\end{array}
\right],
\end{equation}
where the correction matrix to the block-diagonal terms is
\begin{align}
T(\Lambda_+-\Lambda_-)^{-1}T=\left[\begin{array}{cc}
\frac{(t+\delta t_1)^2}{\Delta_v+\delta \Delta_{v,2}} & \frac{(t+\delta t_1)(t+\delta t_2)}{\Delta_v+\delta \Delta_{v,2}}  \\
\frac{(t+\delta t_1)(t+\delta t_2)}{\Delta_v+\delta \Delta_{v,2}} & \frac{(t+\delta t_1)^2}{\Delta_v+\delta \Delta_{v,1}}
+\frac{(t+\delta t_2)^2}{\Delta_v+\delta \Delta_{v,2}}
\end{array}
\right].
\end{align}
\end{widetext}
The upper and lower energy bands, defined by the matrices $\Lambda_+$ and $\Lambda_-$, are repelled by the same value, 
which is defined by the correction matrix $T(\Lambda_+-\Lambda_-)^{-1}T$. Since each block of the full Hamiltonian is $2\times 2$ 
matrix there are two upper bands and two lower bands, as presented in Fig.~\ref{fig:quadrolayer_sample}~(b). 
The calculation of the eigenenergies and eigenstates for the particular case of the quadrolayer Hamiltonian is considered below in 
the subsection \ref{subsec:quadro_excitons}.

\section{Coulomb potential in the S-TMD multilayer}
\label{coulomb}

Let us consider a multilayer of S-TMD encapsulated in between hBN flakes of infinite thickness. 
The surface of the bottom hexagonal boron nitride (hBN) flake has a coordinate $z=0$,
and the surface of the top hBN flake has a coordinate $z=L$. The multilayer crystal is presented as a set 
of $N$ layers of S-TMD, arranged in parallel to $xy$ plane.
The position of $j$-th layer is defined by coordinate $z_j$. We arrange the layers in the following way 
$0<z_1<z_2<\dots z_N<L$.
We suppose that a multilayer can be polarized mainly in the in-plane direction, and weakly in the out-of-plane direction. 
We suppose that an out-of-plane dielectric response of a few layer systems ($N=2,3\dots $) is characterized by 
an out-of-plane dielectric constant $\epsilon_\perp$, which values, however, can be dependent on the number of the layers 
$N$. 

In order to find the potential energy between two charges in S-TMD few-layers we solve the following electrostatic problem. We consider the point-like charge $Q$ at the point $\mathbf{r}=(0,0,z')$, where $0<z'<L$ and calculate electric potential in such a system following \cite{aCudazzo2011}.
Namely, we consider 3 regions: bottom and top hBN semi-infinite media with potentials $\Phi_1(\boldsymbol{\rho},z)$ and $\Phi_3(\boldsymbol{\rho},z)$, and the space between them with the potential $\Phi_2(\boldsymbol{\rho},z)$.
Here we introduced the in-plane vector $\boldsymbol{\rho}=(x,y)$.
The Maxwell equations define the form of these potentials.
Using the cylindrical symmetry of the problem we present the potentials in the form
\begin{equation}
\Phi_j(\boldsymbol{\rho},z)=\frac{1}{(2\pi)^2}\int d^2\mathbf{k} e^{i\mathbf{k}\boldsymbol{\rho}}\Phi_j(\mathbf{k},z)
\end{equation}

The Maxwell's equation $\mathrm{div}\mathbf{D}_{1,3}=0$ for 1-st and 3-rd regions reads
\begin{equation}
-\varepsilon_\parallel\mathbf{k}^2\Phi_{1,3}(\mathbf{k},z)+\varepsilon_\perp\frac{d^2\Phi_{1,3}(\mathbf{k},z)}{dz^2}=0,
\end{equation}
where $\varepsilon_\parallel$ and $\varepsilon_\perp$ are parallel and perpendicular components of the dielectric constant of hBN respectively.
The solutions of these equations are
\begin{align}
\Phi_1(\mathbf{k},z)&=Ae^{\kappa z}, \,\,\,\, \text{for} z\leq 0, \\
\Phi_3(\mathbf{k},z)&=Be^{-\kappa z}, \text{for} z\geq L,
\end{align}
where $\kappa=|\mathbf{k}|\sqrt{\varepsilon_\parallel/\varepsilon_\perp}=
k\sqrt{\varepsilon_\parallel/\varepsilon_\perp}$.

The equation for the potential $\Phi_2(\mathbf{r},z)$ takes a form
\begin{widetext}
\begin{equation}
\Delta_\parallel\Phi(\boldsymbol{\rho},z)+\epsilon_\perp\frac{d^2\Phi(\boldsymbol{\rho},z)}{dz^2}=-4\pi[Q\delta(\boldsymbol{\rho})\delta(z-z') - \varrho_{ind}(\boldsymbol{\rho},z)],
\end{equation}
\end{widetext}
where $\Delta_\parallel$ is 2D Laplace operator and we introduced phenomenological out-of-plane dielectric constant $\epsilon_\perp$.
The first term in square brackets is the charge density of the charge $Q$.
The second term represents the induced charge density due to the polarization of layers by charge $Q$.

$\varrho_{ind}(\boldsymbol{\rho},z)$ is the polarization charge density induced in the planes of the multilayer.
\begin{equation}
\varrho_{ind}(\mathbf{r},z)=\mathrm{div} \mathbf{P}(\boldsymbol{\rho},z).
\end{equation}
Following \cite{aCudazzo2011} we present the polarization in the form
\begin{equation}
\mathbf{P}(\boldsymbol{\rho},z)=\sum_{j=1}^N \delta(z-z_j)\mathbf{P}_\parallel(\boldsymbol{\rho},z_j).
\end{equation}
Using the linear response $\mathbf{P}_\parallel(\boldsymbol{\rho},z_j)=
\chi_{2D}\mathbf{E}_\parallel(\boldsymbol{\rho},z_j)$
we get the expression for the induced charge
\begin{equation}
\varrho_{ind}(\mathbf{r},z)=-\chi_{2D}\sum_{j=1}^N \delta(z-z_j)\Delta_\parallel\Phi_2(\boldsymbol{\rho},z_j).
\end{equation}
Taking Fourier transformation one gets
\begin{widetext}
\begin{align}
\Big[\mathbf{k}^2-\epsilon_\perp\frac{d^2}{dz^2}\Big]\Phi_2(\mathbf{k},z)=4\pi Q\delta(z-z')-
4\pi\chi_{2D}\mathbf{k}^2\sum_{j=1}^N \delta(z-z_j)\Phi_2(\mathbf{k},z_j).
\end{align}
\end{widetext}
Integrating this equation in the regions $z\in[z_j-\epsilon, z_j+\epsilon]$ and $z\in[z'-\epsilon,z'+\epsilon]$ one gets the following conditions
\begin{align}
\epsilon_\perp\frac{d\Phi_2(\mathbf{k},z)}{dz}\Big|^{z_j+\epsilon}_{z_j-\epsilon}&=2r_0k^2\Phi_2(\mathbf{k},z_j),\\
\epsilon_\perp\frac{d\Phi_2(\mathbf{k},z)}{dz}\Big|^{z'+\epsilon}_{z'-\epsilon}&=-4\pi Q,
\end{align}
where we introduced $r_0=2\pi\chi_{2D}$. Outside of the the points $z_j$ and $z'$ we have the equation
\begin{equation}
\Big[\mathbf{k}^2-\epsilon_\perp\frac{d^2}{dz^2}\Big]\Phi_2(\mathbf{k},z)=0
\end{equation}
The general solution for $\Phi_2(\mathbf{k},z)$ can be written in the form
\begin{widetext}
\begin{equation}
\label{eq:potential2}
\Phi_2(\mathbf{k},z)=\Psi_0e^{-K|z-z'|}+\sum_{j=1}^N \Psi_je^{-K|z-z_j|} + \alpha e^{Kz} +\beta e^{-Kz},
\end{equation}
\end{widetext}
where $\Psi_0$,$\Psi_j$, $\alpha$ and $\beta$ are unknown functions of $K=k/\sqrt{\epsilon_\perp}$.
The aforementioned boundary conditions together with the equation give the following restrictions
\begin{widetext}
\begin{align}
\label{eq:boundary_conditions}
\Psi_j=-r_0k\Phi_2(\mathbf{k},z_j)/\sqrt{\epsilon_\perp}=-r_0K\Phi_2(\mathbf{k},z_j), \quad
\Psi_0=2\pi Q/k\sqrt{\epsilon_\perp}=2\pi Q/\epsilon_\perp K.
\end{align}
\end{widetext}
Taking the boundary conditions at $z=0$ and $z=L$
\begin{widetext}
\begin{align}
\Phi_1(\mathbf{k},0)=\Phi_2(\mathbf{k},0), \quad  \frac{\varepsilon_\perp}{\epsilon_\perp}
\frac{d\Phi_1(\mathbf{k},z)}{dz}\Big|_{z=0}=\frac{d\Phi_2(\mathbf{k},z)}{dz}\Big|_{z=0},  \\
\Phi_3(\mathbf{k},L)=\Phi_2(\mathbf{k},L), \quad \frac{\varepsilon_\perp}{\epsilon_\perp}
\frac{d\Phi_3(\mathbf{k},z)}{dz}\Big|_{z=L}=\frac{d\Phi_2(\mathbf{k},z)}{dz}\Big|_{z=L},
\end{align}
\end{widetext}
we get the following equations
\begin{widetext}
\begin{align}
&A=\Psi_0e^{-Kz'}+\sum_{j=1}^N\Psi_je^{-Kz_j}+\alpha+\beta, \quad 
\varepsilon A=\Psi_0e^{-Kz'}+\sum_{j=1}^N\Psi_je^{-Kz_j}+\alpha-\beta, \\
&Be^{-\kappa L}=\Psi_0e^{-K(L-z')}+\sum_{j=1}^N\Psi_je^{-K(L-z_j)}+\alpha e^{KL}+\beta e^{-KL},  \\
&\varepsilon Be^{-\kappa L}=\Psi_0e^{-K(L-z')}+\sum_{j=1}^N\Psi_je^{-K(L-z_j)}-\alpha e^{KL}+\beta e^{-KL}, 
\end{align}
\end{widetext}
where we introduced $\varepsilon=\sqrt{\varepsilon_\perp\varepsilon_\parallel/\epsilon_\perp}$. 
Removing parameters $A$ and $B$ from these equations one gets for $\varepsilon\neq1$
\begin{align}
\Psi_0e^{-Kz'}+\sum_{j=1}^N\Psi_je^{-Kz_j}+\alpha+\beta/\delta=0,
\end{align}
\begin{align}
\Psi_0e^{-K(L-z')}+\sum_{j=1}^N\Psi_je^{-K(L-z_j)}+\alpha e^{KL}/\delta+\beta e^{-KL}=0, 
\end{align}
where $\delta=(\varepsilon-1)/(\varepsilon+1)\in(0,1)$. 
Note that for the case $\varepsilon=1$ we get $\alpha=\beta=0$, and therefore the solution in the second region
\begin{equation}
\Phi_2(\mathbf{k},z)=\Psi_0e^{-K|z-z'|}+\sum_{j=1}^N \Psi_je^{-K|z-z_j|},
\end{equation}

We solve the general equation in a few steps. First, we write the system of equations in the following form
\begin{align}
\label{eq:alphabeta}
\delta\alpha+\beta=\delta\mathcal{A}, \quad e^{KL}\alpha+\delta e^{-KL}\beta=\delta\mathcal{B},
\end{align}
where we introduced $\mathcal{A}=-\Psi_0e^{-Kz'}-\sum_{j=1}^N\Psi_je^{-Kz_j}$, $\mathcal{B}=-\Psi_0e^{-K(L-z')}-\sum_{j=1}^N\Psi_je^{-K(L-z_j)}$.
Solving the equations (\ref{eq:alphabeta}) one gets
\begin{align}
\label{eq:alphabeta}
\alpha=\frac{\delta[\mathcal{B}e^{KL}-\mathcal{A}\delta]}{e^{2KL}-\delta^2}, \quad \beta=\frac{\delta e^{KL}[\mathcal{A}e^{KL}-\mathcal{B}\delta]}{e^{2KL}-\delta^2}.
\end{align}
Substituting these solutions in (\ref{eq:potential2}) we express $\Phi_2(\mathbf{k},z)$ via $N+1$ parameters $\Psi_0$ and $\Psi_j$. Then, using the latter expression,
we calculate the expressions for the potential in $z'$ and $z_l,\, l=1,2\dots N$ points
\begin{widetext}
\begin{equation}
\label{eq:self_consisted}
\Phi_2(\mathbf{k},z_l)=\Psi_0e^{-K|z_l-z'|}+\sum_{j=1}^N \Psi_je^{-K|z_l-z_j|} + \alpha e^{Kz_l} +\beta e^{-Kz_l},
\end{equation}
\end{widetext}
which together with (\ref{eq:boundary_conditions}) gives the complete set of equations for parameters 
$\Psi_0$ and $\Psi_j$.

Note that the system of equations and (\ref{eq:boundary_conditions}), (\ref{eq:alphabeta}) and 
(\ref{eq:self_consisted}) are simplified at 
$\epsilon_\perp=1$. We obtain the solution for this particular case first for the purpose of brevity. 
The general solution for arbitrary $\epsilon_\perp$ can be obtained from the previous solution 
by replacing $k\rightarrow K=k/\sqrt{\epsilon_\perp}$, $Q\rightarrow Q/\epsilon_\perp$, and 
$\varepsilon\rightarrow\sqrt{\varepsilon_\perp\varepsilon_\parallel/\epsilon_\perp} $ as one can see from the aforementioned system of equations and definition of $\varepsilon$ and $K$.

\subsection{Monolayer}

We solve the system of equations (\ref{eq:self_consisted}) for the S-TMD monolayer encapsulated in between two dielectric flakes with distance $L$ between them.
We consider the case when the charge $Q$ is placed inside the monolayer $z'=z_1$. We consider the case of a symmetric disposition of the monolayer between dielectric media $z_1=L/2$.
The solution is
\begin{equation}
\Phi_2(\mathbf{k},L/2)=\frac{2\pi Q}{k}\frac{1-\delta e^{-kL}}{1+kr_0-\delta(kr_0-1)e^{-kL}}.
\end{equation}
In $L\rightarrow\infty$ limit the answer doesn't depend on $\delta$ and reproduces the case of the potential for the monolayer in vacuum
\begin{equation}
\Phi_2(\mathbf{k},L/2)=\frac{2\pi Q}{k}\frac{1}{1+kr_0}.
\end{equation}
At $L\rightarrow0$ limit it coincides with the Rytova-Keldysh potential \cite{aCudazzo2011,aKeldysh1979}
\begin{align}
\Phi_2(\mathbf{k},0)=\frac{2\pi Q}{k}\frac{1-\delta}{1+\delta+(1-\delta)kr_0}=
\frac{2\pi Q}{k\varepsilon}\frac{1}{1+kr_0/\varepsilon}.
\end{align}
Using Fourier transformation of the latter expression we restore the Rytova-Keldysh potential
\begin{equation}
V_{RK}(\rho)=%\frac{Q}{r_0}\int_0^\infty dx\frac{J_0\Big(x\frac{\rho\varepsilon}{r_0}\Big)}{1+x}=
\frac{\pi Q}{2r_0}\Big[\text{H}_0\Big(\frac{\rho\varepsilon}{r_0}\Big)-Y_0\Big(\frac{\rho\varepsilon}{r_0}\Big)\Big].
\end{equation}

\subsection{Bilayer}

We consider two monolayers positioned at $z_1=\xi$ and $z_2=L-\xi$. We calculate the in-plane potential 
$\Phi_2(\mathbf{k},z_1)$ with the charge
$Q$ placed in $z'=z_1$ and  then consider the limit $\xi\rightarrow0$ for simplicity. The answer is
\begin{widetext}
\begin{align}
\Phi_2(\mathbf{k},0)=&\frac{2\pi Q}{k}\frac{(1-\delta)\left(e^{2kL}[(1-\delta)kr_0+1]-[kr_0(1-\delta)+\delta]\right)}
{e^{2kL}[(1-\delta)kr_0+1]^2-[\delta+kr_0(1-\delta)]^2},\\
\Phi_2(\mathbf{k},L)=&\frac{2\pi Q}{k}\frac{(1-\delta)^2e^{kL}}{e^{2kL} [(1-\delta)kr_0+1]^2-[\delta +kr_0(1-\delta)]^2}.
\end{align}
\end{widetext}
The limit case $L\rightarrow0$ gives
\begin{align}
\Phi_2(\mathbf{k},0)=\Phi_2(\mathbf{k},L\rightarrow 0)=
\frac{2\pi Q}{k\varepsilon}\frac{1}{1+2kr_0/\varepsilon}.
\end{align}
For $L\rightarrow\infty$ we have $\Phi_2(\mathbf{k},L\rightarrow\infty)=0$
\begin{align}
\Phi_2(\mathbf{k},0)=\frac{2\pi Q}{k\varepsilon_{eff}}\frac{1}{1+kr_0/\varepsilon_{eff}}.
\end{align}
Here $\varepsilon_{eff}=(\varepsilon+1)/2$.

\subsection{Trilayer}
\label{sec:trilayer}

We consider three monolayers positioned at $z_1=\xi$, $z_2=L/2$ and $z_3=L-\xi$. We calculate the in-plane potentials for two cases,
when the charge $Q$ is placed in first plane $z'=z_1$ and when the charge $Q$ is placed in the second plane $z'=z_2$. 
The potential when the charge is localized in the top layer with $z'=z_3$ can be restored from the case $z'=z_1$ using 
symmetry arguments. 
We consider
the limit $\xi\rightarrow0$ for simplicity. Due to the symmetry of the system, the situation with charge in the top layer can be restored from the first case. 

\subsubsection{Charge is placed in the bottom plane}

We present the potential in $j(=1,2,3)$th layer, induced by the charge placed in the bottom (first) layer as
\begin{align}
\Phi_2(\mathbf{k},z_j)=\frac{2\pi Q}{k}\frac{(1-\delta)h_{1j}(k)}{p(k)},
\end{align} 
where 
\begin{widetext}
\begin{align}
h_{11}(k)=&\big[(1+kr_0)(1+[1-\delta]kr_0)-kr_0(1+2kr_0[1-\delta]+\delta)e^{-kL}+(kr_0-1)(kr_0[1-\delta]+\delta)e^{-2kL}\big], \\
h_{12}(k)=&e^{-kL/2}\big[(1 + kr_0[1 -\delta]) - (\delta + kr_0[1-\delta])e^{-k L}\big], \\
h_{13}(k)=&(1-\delta)e^{-kL}, \\
p(k)=&
{(1+kr_0)(1+[1-\delta]kr_0)^2-2kr_0e^{-kL}(1+kr_0[1-\delta])(kr_0[1-\delta]+\delta)+
e^{-2kL}(kr_0-1)(kr_0[1-\delta]+\delta)^2}= \nonumber \\ 
=&\big[(1 + kr_0[1 -\delta]) - (\delta + kr_0[1-\delta])e^{-k L}\big]
\big[(kr_0+1)(kr_0[1-\delta]+1)-e^{-k L} (kr_0-1) (\delta +kr_0[1-\delta])\big] .
\end{align}
\end{widetext}
%The limit case $L\rightarrow0$ is
%\begin{align}
%\Phi_2(\mathbf{k},0)=\frac{2\pi Q}{k\varepsilon}\frac{1}{1+3kr_0/\varepsilon},
%\end{align}
%while for $L\rightarrow\infty$ is
%\begin{align}
%\Phi_2(\mathbf{k},0)=\frac{2\pi Q}{k\varepsilon_{eff}}\frac{1}{1+kr_0/\varepsilon_{eff}}.
%\end{align} 
%The potential in the second layer is 
%\begin{widetext}
%\begin{equation}
%\label{eq:v12}
%\Phi_2(\mathbf{k},L/2)=\frac{2\pi Q}{k}\frac{(1-\delta)e^{-kL/2}}{(kr_0+1)(kr_0[1-\delta]+1)-
%e^{-kL}(kr_0-1)(kr_0[1-\delta]+\delta)}.
%\end{equation}
%The potential in the third layer is 
%\begin{equation}
%\Phi_2(\mathbf{k},L)=\frac{2\pi Q}{k} \frac{(1-\delta)^2e^{-kL}}{(1+kr_0)(1+[1-\delta]kr_0)^2-2kr_0e^{-kL}
%(1+kr_0[1-\delta])(kr_0[1-\delta]+\delta)+e^{-2kL}(kr_0-1)(kr_0(1-\delta)+\delta)^2}.
%\end{equation}

\subsubsection{Charge is placed in the middle plane}

For the second case, one needs to find the potential in the second plane $z_2=L/2$, induced by the charge $Q$, placed in the same plane. The potentials for the first $z_1=0$ and third $z_3=L$ planes can be obtained from the previous 
case using the symmetry properties of the system. The potential has a form  
\begin{align}
\Phi_2(\mathbf{k},L/2)=\frac{2\pi Q}{k}\frac{h_{22}(k)}{p(k)},
\end{align} 
where we introduced 
\begin{align}
h_{22}(k)=\big[(1+[1-\delta]kr_0)-(kr_0[1-\delta]+\delta)e^{-kL}\big]^2.
\end{align}
%The limit case $L\rightarrow0$ is
%\begin{align}
%\Phi_2(\mathbf{k},0)=\frac{2\pi Q}{k\varepsilon}\frac{1}{1+3kr_0/\varepsilon},
%\end{align}
%while for $L\rightarrow\infty$ is
%\begin{align}
%\Phi_2(\mathbf{k},0)=\frac{2\pi Q}{k}\frac{1}{1+kr_0}.
%\end{align}
%\begin{widetext}
%\begin{equation}
%\Phi_2(\mathbf{k},0)=\Phi_2(\mathbf{k},L)=\frac{2\pi Q}{k}\frac{(1-\delta)e^{-kL/2}}{(kr_0+1)([1-\delta]kr_0+1)-
%e^{-kL}(kr_0-1)(\delta+[1-\delta]kr_0)}.
%\end{equation}
%\end{widetext}
%Note that this result coincides with the expression (\ref{eq:v12}) from the previous subsection, as it should be 
%from the symmetry reasons.  

\subsection{Quadrolayer}
\label{sec:quadrolayer}

We consider four monolayers positioned at $z_1=\xi$, $z_2=L/3$, $z_3=2L/3$ and $z_4=L-\xi$. We calculate the in-plane potentials for two cases, for the charge $Q$ is placed in the first plane $z'=z_1$  and for the charge $Q$ is placed in the second plane $z'=z_2$. For the first case, we calculate the potentials in all layers separately. For the second case we calculate the potentials in the second and the third layers. All other potentials can be restored from the former ones by applying the symmetry arguments for the considered system. 
We consider the limit $\xi\rightarrow0$ for simplicity. 

\subsubsection{Charge placed in the bottom plane}

We present the potential in $j(=1,2,3,4)$-th layer, induced by the charge $Q$, placed in the first layer, as 
\begin{equation}
\Phi_2(\mathbf{k},z_j)=\frac{2\pi Q}{k}\frac{(1-\delta)f_{1j}(k)}{g(k)},
\end{equation}
Here 
\begin{widetext}
\begin{align}
f_{11}(k)=&(1+kr_0)^2(1+[1-\delta]kr_0)-kr_0(1+kr_0[4-\delta]+3k^2r_0^2[1-\delta]+\delta)e^{-2kL/3}- \nonumber \\
&-kr_0(1-3k^2r_0^2[1-\delta]+\delta+kr_0[1-4\delta])e^{-4kL/3}-(kr_0-1)^2(kr_0[1-\delta]+\delta)e^{-2kL},\\
f_{12}(k)=&(kr_0+1)(1+kr_0[1-\delta])e^{-kL/3}-kr_0(1+\delta +2kr_0[1-\delta])e^{-kL}+
(kr_0-1)(\delta+kr_0[1-\delta])e^{-5kL/3}, \\
f_{13}(k)=&(1+kr_0[1-\delta])e^{-2kL/3}-(\delta+kr_0[1-\delta])e^{-4kL/3}, \\
f_{14}(k)=&(1-\delta)e^{-kL}, \\
g(k)=&(1+kr_0)^2(1+[1-\delta]kr_0)^2-kr_0(kr_0[1-\delta]+1)(3kr_0[1+kr_0(1-\delta)]+2\delta)e^{-2kL/3}+ \nonumber \\
&+kr_0(-2+3kr_0[kr_0(1-\delta)+\delta])(kr_0[1-\delta]+\delta)e^{-4kL/3}-(kr_0-1)^2(kr_0[1-\delta]+\delta)^2e^{-2kL}. 
\end{align}

\end{widetext}

\subsubsection{Charge placed in the second plane}

Similarly to the previous case we present the potential in $j(=1,2,3,4)$-th layer, induced by the charge $Q$, 
placed in the second layer, as 
\begin{equation}
\Phi_2(\mathbf{k},z_j)=\frac{2\pi Q}{k}\frac{f_{2j}(k)}{g(k)}.
\end{equation} 
Taking into account the symmetry of the system, it is enough to derive $f_{22}(k)$ and $f_{23}(k)$. 
The calculation gives the following 
\begin{widetext}
\begin{align}
f_{22}(k)=&\big[(1+kr_0)(1+kr_0[1-\delta])-kr_0(1+2kr_0[1-\delta]+\delta)e^{-2kL/3}+(kr_0-1)(kr_0[1-\delta]+\delta)e^{-4kL/3}\big]\times \nonumber \\
&\times\big[(1+kr_0[1-\delta])-(kr_0[1-\delta]+\delta)e^{-2kL/3}\big], \\
f_{23}(k)=&e^{-k L/3}\Big[(1+kr_0[1-\delta])-e^{-2kL/3}(\delta+kr_0[1-\delta])\Big]^2.
\end{align}
\end{widetext}
%The limit case $L\rightarrow0$ is
%\begin{align}
%\Phi_2(\mathbf{k},L/3)=\frac{2\pi Q}{k\varepsilon}\frac{1}{1+4kr_0/\varepsilon},
%\end{align}
%while for $L\rightarrow\infty$ is
%\begin{align}
%\Phi_2(\mathbf{k},L/3)=\frac{2\pi Q}{k}\frac{1}{1+kr_0}.
%\end{align}

\section{Effective Hamiltonian for intralayer excitons in few-layers of WSe$_2$}
\label{effective}

The binding energies of excitons in monolayer are investigated in \cite{aMolas2019}.    
To investigate the case of bi-, tri-, and quadrolayer of transition metal dichalcogenides
we consider $\mathbf{k\cdot p}$ approximation and derive the corresponding effective Hamiltonian 
for the intravalley excitons \cite{aKormanyos2015,aSlobodeniuk2019}. 
We start from the case $\epsilon_\perp=1$, as one of the limit cases of our model. 

\subsection{Monolayer} 

The spectrum of excitons in the S-TMD monolayer can be obtained by solving the equation
\begin{equation}
\label{eq:eff_two_body_hamiltonian}
\Big\{\frac{\hbar^2}{2\mu}\nabla_\parallel^2-V_{RK}(\rho)+E\Big\}\psi(\rho)=0, 
\end{equation} 
where  
\begin{equation}
V_{RK}=-\frac{\pi e^2}{2r_0}\Big[\text{H}_0\Big(\frac{\rho\varepsilon}{r_0}\Big)-
Y_0\Big(\frac{\rho\varepsilon}{r_0}\Big)\Big].
\end{equation}
Introducing the dimensionless parameters $\xi=\rho\varepsilon/r_0=\rho/r_0^*$ and $E=(\mu e^4/2\hbar^2\varepsilon^2)\epsilon=Ry^*\epsilon$ one rewrites the equation for the case of s-type excitons in the following dimensionless form
\begin{equation}
\Big\{b^2\frac{1}{\xi}\frac{d}{d\xi}\Big(\xi\frac{d}{d\xi}\Big)+bv_{RK}(\xi)+\epsilon\Big\}\psi(\xi)=0,
\end{equation} 
where $b=\hbar^2\varepsilon^2/\mu e^2r_0$ and 
\begin{equation}
v_{RK}(\xi)=\pi[\text{H}_0(\xi)-Y_0(\xi)].
\end{equation}
For WSe$_2$ monolayer $r_0=45\mbox{\AA}$, $\mu=0.21m_0$, 
where $m_0$ is electron's mass \cite{aKormanyos2015,aMolas2019,aBerkelbach2013}. 
We use $\varepsilon=4.5$ for the dielectric constant of hBN. Then $b\approx 1.136$. 
Taking the numerical solution of the equation for this value of $b$ gives the value
$Ry^*\approx 141$\,meV  we obtain the following binding energies of the excitons:
$E_1\approx-165$\,meV,  $E_2\approx-39$\,meV, $E_3\approx -17 $\,meV, and $E_4\approx -9 $\,meV,.
 
In order to compare numerical and experimental results we consider the difference between neighbor 
energies of the spectrum $\Delta E_j=E_j-E_1$: 
$\Delta E_2 \approx 126  $~meV, $\Delta E_3 \approx 148 $~meV, $\Delta E_4 \approx 156$~meV.

\subsection{Bilayer} 

\begin{figure}[t]
	\centering
	\includegraphics[width=0.75\linewidth]{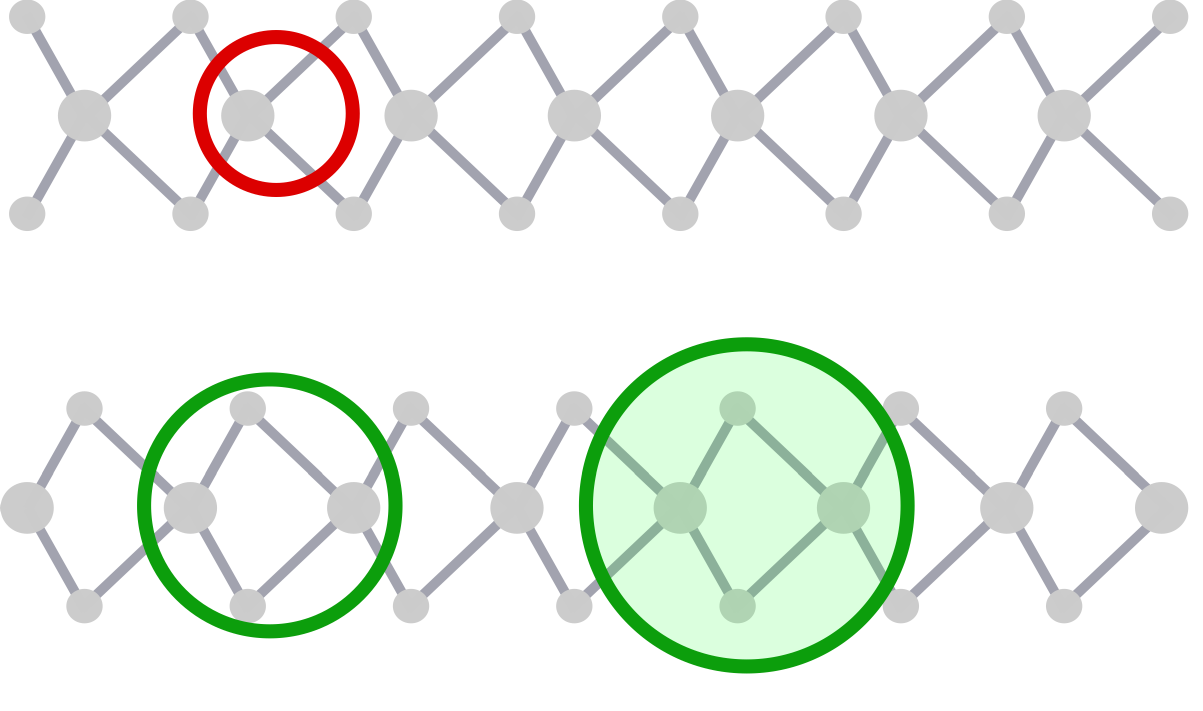}%
	\caption{The schematic picture of the intralayer exciton in the bilayer S-TMD in K$^+$ point. 
	The filled circle represents the conduction band electron localized in the bottom layer. 
	Two empty circles represent the hole quasiparticle in the bilayer, localized mainly in the bottom layer. 
	The size of the circles denotes the amount of charge of quasiparticles redistributed in each layer. 
	The colors of the circles encode the polarizations of photons which can be coupled to the exciton: 
	green for $\sigma^+$ and red for $\sigma^-$. The latter means that the intralayer exciton of the bilayer 
	in K$^+$ point is mainly active in $\sigma^+$ polarization and much smaller in $\sigma^-$ polarization.} 
	\label{fig:bilayer_2022}
\end{figure}

To calculate the binding energies of A-excitons in the bilayer one needs to take into account the admixture of the
valence band states from the opposite valleys in different layers. Due to this admixture, the charge of a hole is redistributed between two layers, see Fig.~\ref{fig:bilayer_2022}.
It modifies the Coulomb potential between electrons and holes in the bilayer. 
Taking into account the results of Ref.~\cite{aSlobodeniuk2019} we obtain the following 
value for charges of the hole, which belongs to the same plane as an electron 
\begin{equation}
Q_\textrm{in}=\frac{|e|}{2}\Big(1+\frac{\Delta_v}{\sqrt{\Delta_v^2+4t^2}}\Big)\approx 0.98|e|,
\end{equation}
and for the charge of the hole, which belongs to the opposite plane as an electron
\begin{equation}
Q_\textrm{opp}=\frac{|e|}{2}\Big(1-\frac{\Delta_v}{\sqrt{\Delta_v^2+4t^2}}\Big)\approx 0.02|e|.
\end{equation} 
Here we used the numbers $\Delta_v=456$\,meV, $t=67$\,meV from Ref.~\cite{aGong2013}.
As one can see, most of the charge of the hole is localized within one layer.
It means that the binding energy of the intralayer exciton in the bilayer should be close 
to the ones in the monolayer. 

Using the values for charges and the expressions for the potentials in the bilayer we present
the Coulomb potential between electron and a hole as
\begin{widetext}
\begin{align}
V_\text{bil}(k)=-\frac{2\pi e^2}{k\sqrt{\epsilon_\perp}} 
\frac{0.98(1-\widetilde{\delta})\left(e^{2k\widetilde{L}}[(1-\widetilde{\delta})k\widetilde{r_0}+1]-[k\widetilde{r_0}(1-\widetilde{\delta})+\widetilde{\delta}]\right)
+0.02(1-\widetilde{\delta})^2e^{k\widetilde{L}}}
{e^{2k\widetilde{L}}[(1-\widetilde{\delta})k\widetilde{r_0}+1]^2-[\widetilde{\delta}+k\widetilde{r_0}
(1-\widetilde{\delta})]^2}.
\end{align}
\end{widetext}
Here the parameters $\widetilde{\delta}=(\varepsilon-\sqrt{\epsilon_\perp})/(\varepsilon+\sqrt{\epsilon_\perp})
\approx 0.24$, $\epsilon_\perp=7.6$ \cite{aLaturia2020}, 
$\widetilde{L}=L/\sqrt{\epsilon_\perp}\approx 2.354\,\mbox{\AA}$ (HQ Graphene) and $\widetilde{r_0}=r_0/\sqrt{\epsilon_\perp}=16.323\,\mbox{\AA}$ \cite{aBerkelbach2013}.

The coordinate-dependent potential then reads
\begin{widetext}
\begin{align}
V_\text{bil}(\rho)=-\frac{e^2}{\sqrt{\epsilon_\perp}}\int_0^\infty dk J_0(k\rho)
\frac{0.98(1-\widetilde{\delta})\left(e^{2k\widetilde{L}}[(1-\widetilde{\delta})k\widetilde{r_0}+1]-[k\widetilde{r_0}(1-\widetilde{\delta})+\widetilde{\delta}]\right)
+0.02(1-\widetilde{\delta})^2e^{k\widetilde{L}}}
{e^{2k\widetilde{L}}[(1-\widetilde{\delta})k\widetilde{r_0}+1]^2-[\widetilde{\delta}+k\widetilde{r_0}
(1-\widetilde{\delta})]^2}
\end{align}
\end{widetext}  
The effective masses of the top $m_\text{vb}^\text{top}=0.36m_0$ and bottom 
$m_\text{vb}^\text{bott}=0.54m_0$ valence bands in WSe$_2$ are not the same 
\cite{aKormanyos2015}. Here $m_0$ is an electron mass.
Therefore, the effective mass of the new hole quasiparticles is a weighted value of the 
the top and bottom masses. However, since the main part of the hole quasiparticle is mainly localized in the
one layer assume the reduced mass as in monolayer $\mu=0.21m_0$ as a first approximation.

The energies of 1s and 2s intralayer excitons in the bilayer can be found from the equation 
\begin{equation}
\Big\{\frac{\hbar^2}{2\mu}\nabla_\parallel^2-V_\text{bil}(\rho)+E\Big\}\psi(\rho)=0.
\end{equation}
Introducing again the dimensionless parameters of energy $\epsilon=E/(\mu e^2/2\hbar^2\varepsilon^2)$ and 
distance $\xi=\rho\varepsilon/\widetilde{r_0}$ we obtain the following equation
\begin{equation}
\Big\{b^2\epsilon_\perp\frac{1}{\xi}\frac{d}{d\xi}\Big(\xi\frac{d}{d\xi}\Big)+
bv_\text{bil}(\xi)+\epsilon\Big\}\psi(\xi)=0,
\end{equation} 
where 
\begin{widetext}
\begin{align}
\label{eq:v_bil}
v_\text{bil}(\xi)=2\varepsilon\int_0^\infty dx J_0(x\xi)
\frac{0.98(1-\widetilde{\delta})\left(e^{2xl}[(1-\widetilde{\delta})\varepsilon x+1]-[\varepsilon x(1-\widetilde{\delta})+\widetilde{\delta}]\right)
+0.02(1-\widetilde{\delta})^2e^{xl}}
{e^{2xl}[(1-\widetilde{\delta})\varepsilon x+1]^2-[\widetilde{\delta}+\varepsilon x(1-\widetilde{\delta})]^2}
\end{align}
\end{widetext}
and $l=L\varepsilon/r_0\approx 0.649$.   

Solving numerically the eiegenvalue problem we get $E_1\approx -140$\,meV, $E_2\approx -35 $\, meV, 
$E_3\approx -16 $\,meV, $E_4\approx-9 $\, meV.
The differences of the binding energies $\Delta E_j=E_j-E_1$ then are $\Delta E_2\approx 105$\,meV, 
$\Delta E_3\approx 124$\,meV, $\Delta E_4\approx 131$\,meV.  

\subsection{Trilayer} 

\begin{figure}[t]
	\centering
	\includegraphics[width=0.75\linewidth]{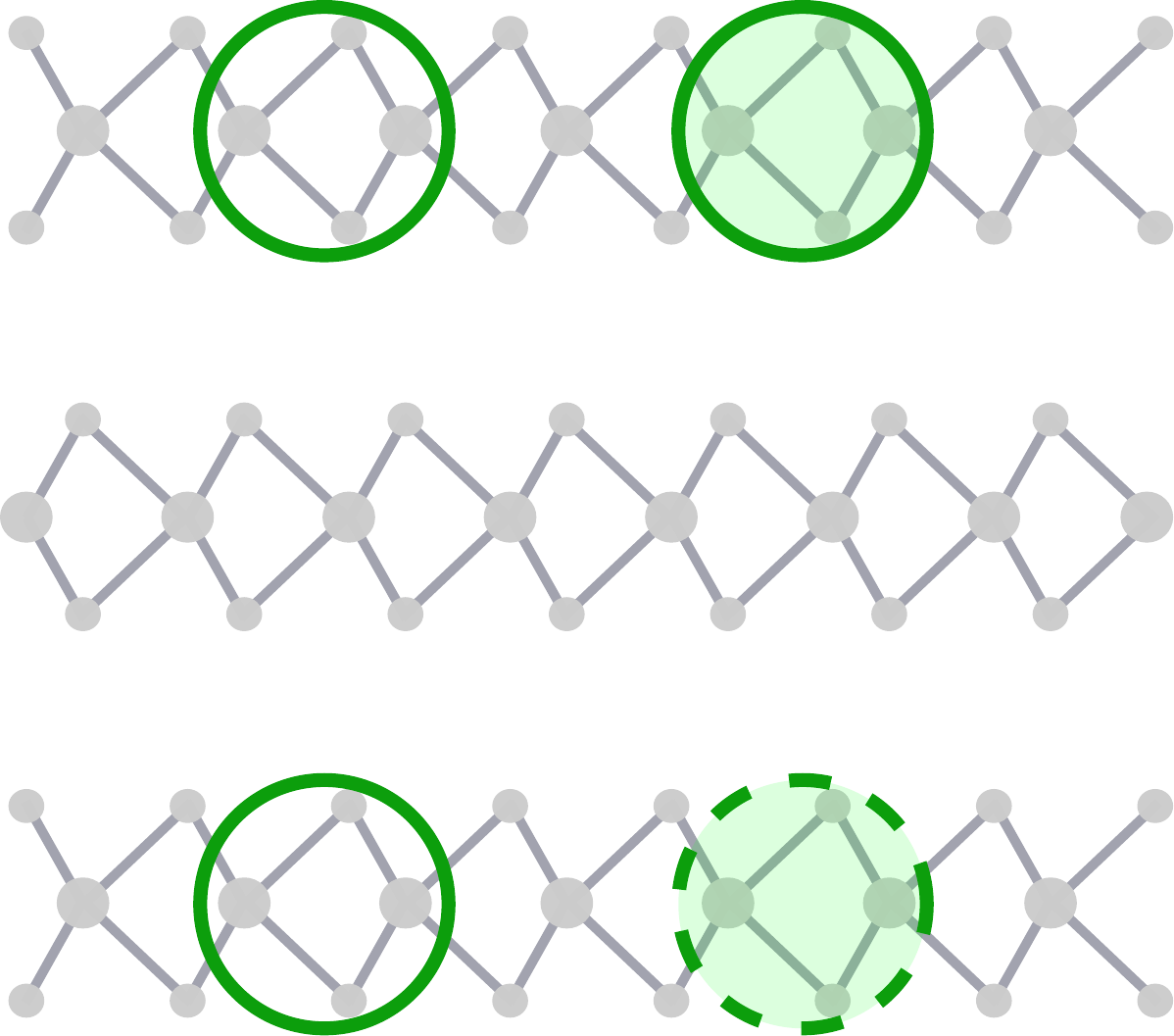}%
	\caption{The schematic picture of the intralayer antisymmetric exciton in the trilayer S-TMD in K$^+$ point. 
	The filled circles represent the conduction band electron quasiparticle localized either in the top (solid circle) 
        or in the bottom (dashed circle) layers. 
	Two empty circles represent the hole quasiparticle, localized in the top and bottom layers. 
	The size of the circles denotes the amount of charge of quasiparticles redistributed in each layer. 
	Due to the symmetry of the crystal the top and bottom layers carry the same amount of charge.
	The colors of the circles encode the polarizations of photons which can be coupled to the exciton: 
	green for $\sigma^+$ and red for $\sigma^-$. The latter means that the antisymmetric exciton of trilayer 
	in K$^+$ point is only active in $\sigma^+$ polarization.} 
	\label{fig:trilayer_2022_odd}
\end{figure}

\begin{figure}[t]
	\centering
	\includegraphics[width=0.75\linewidth]{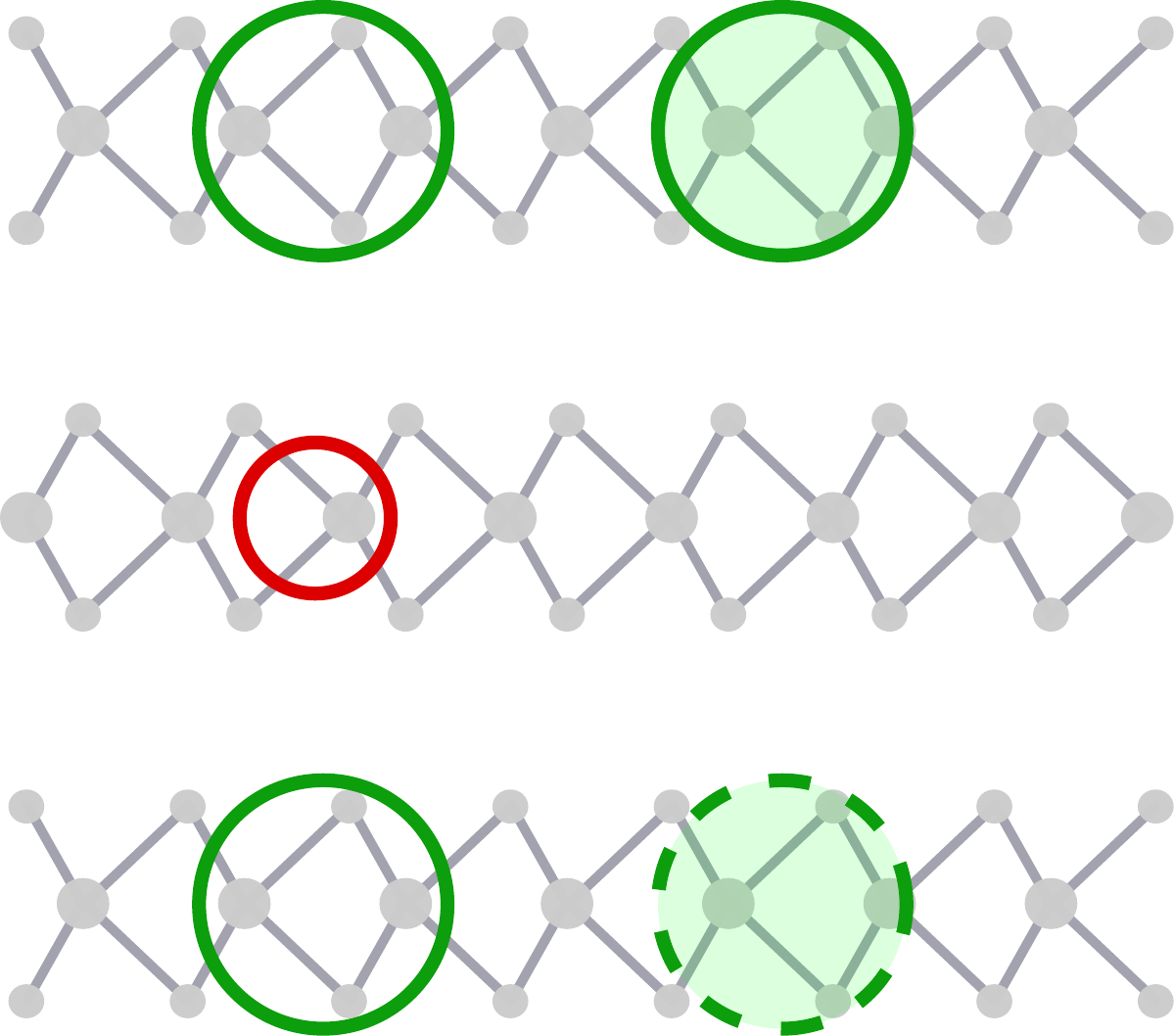}%
	\caption{The schematic picture of the intralayer symmetric exciton in the trilayer S-TMD in K$^+$ point. 
	The filled circles represent the conduction band electron quasiparticle localized either in the top (solid circle) 
        or in the bottom (dashed circle) layers. 
	The empty circles represent the hole quasiparticle in the trilayer, localized mainly in the top and bottom layers. 
	The size of the circles denotes the amount of charge of quasiparticles redistributed in each layer. 
	The colors of the circles encode the polarizations of photons which can be coupled to the exciton: 
	green for $\sigma^+$ and red for $\sigma^-$. The latter means that the corresponding intralayer exciton   
	in K$^+$ point is mainly active in $\sigma^+$ polarization and much smaller in $\sigma^-$ polarization.} 
	\label{fig:trilayer_2022_even_1}
\end{figure}

\begin{figure}[t]
	\centering
	\includegraphics[width=0.75\linewidth]{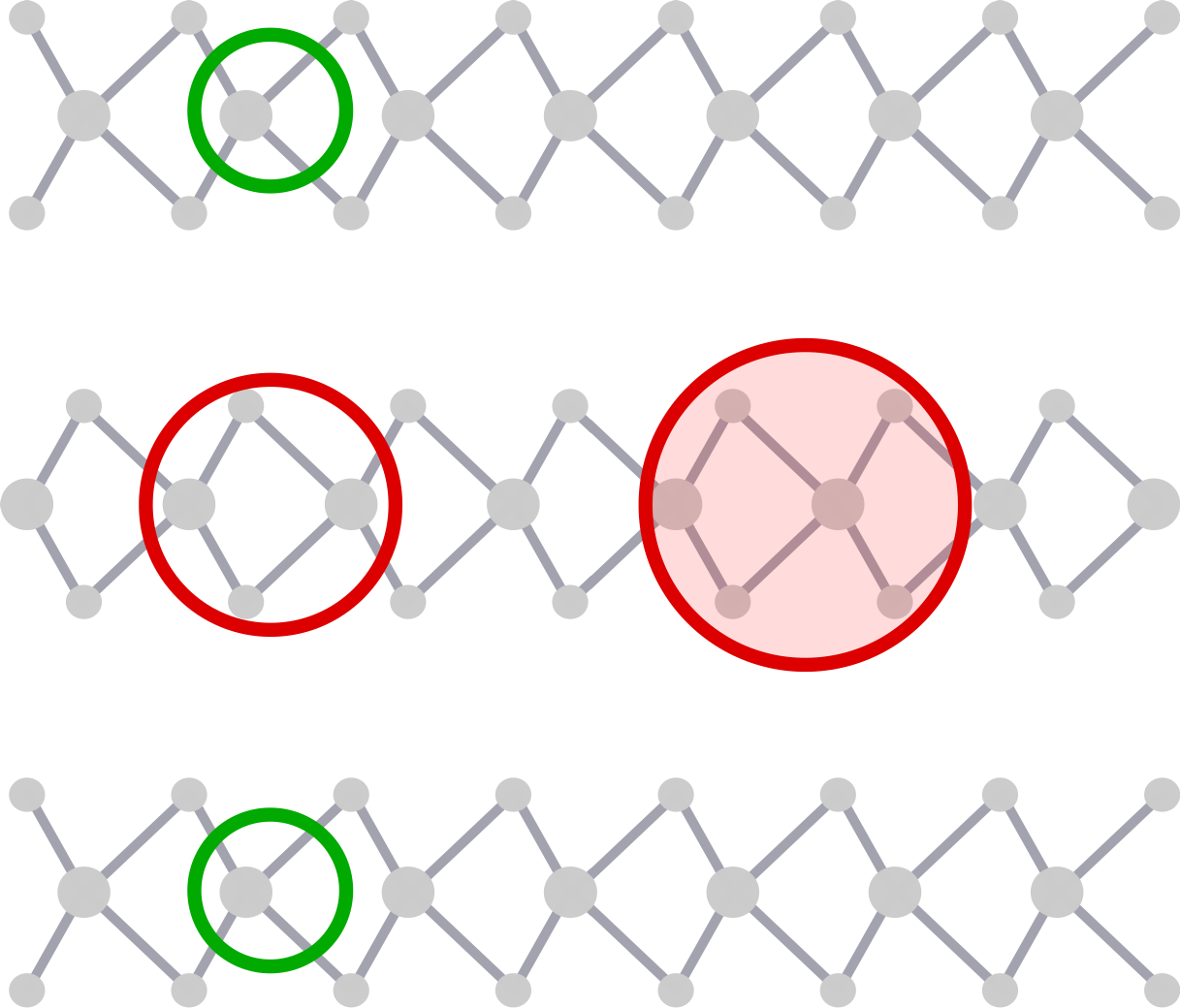}%
	\caption{The schematic picture of another intralayer symmetric exciton in the trilayer S-TMD in K$^+$ point. 
	The filled circle represents the conduction band electron localized in the middle layer. 
	The empty circles represent the hole quasiparticle in the trilayer, localized mainly in the middle layer. 
	The size of the circles denotes the amount of charge of quasiparticles redistributed in each layer. 
	The colors of the circles encode the polarizations of photons which can be coupled to the exciton: 
	green for $\sigma^+$ and red for $\sigma^-$. The latter means that this exciton  
	in K$^+$ point is mainly active in $\sigma^-$ polarization and much smaller in $\sigma^+$ polarization.} 
	\label{fig:trilayer_2022_even_2}
\end{figure}

The structure of the excitonic states in the trilayer is more complex than in the bilayer. 
There are three types of conduction band electrons: localized in the bottom, middle, and top layer, respectively. 
There are three types of hole quasiparticles: antisymmetric superposition of the hole excitations 
in the top and bottom layers, two orthogonal to each other symmetrical superpositions of the hole excitations from all the layers of the crystal. These electron and hole quasiparticles form antisymmetrical exciton states,  
see Fig.~\ref{fig:trilayer_2022_odd}, and two symmetrical types of excitons, represented in Figs.~\ref{fig:trilayer_2022_even_1} and \ref{fig:trilayer_2022_even_2}.
We consider these three types of excitons separately. 

\subsubsection{Antisymmetric excitons}

Note that the effective masses of the electron and hole quasiparticles for the antisymmetric combinations of electron and holes are the same as in monolayer \cite{aKormanyos2015}. Therefore the binding energies of the corresponding excitons are defined by the Coulomb interaction between the corresponding charge complexes in the trilayer.   
In the case of the antisymmetric exciton, the top and bottom charges of hole quasiparticles are half of the electron charge (by absolute value). Taking into account the results of the Sec.~\ref{sec:trilayer} we obtain the following expression for the potential between the electron and hole quasiparticle complexes 
\begin{equation}
V_\textrm{tri}^\textrm{as}(k)=-\frac{\pi e^2}{k\sqrt{\epsilon_\perp}}\frac{(1-\widetilde{\delta})}
{\widetilde{p}(k)}\big[\widetilde{h}_{11}(k)+\widetilde{h}_{13}(k)\big], 
\end{equation}
where $\widetilde{\delta}=(\varepsilon-\sqrt{\epsilon_\perp})/(\varepsilon-\sqrt{\epsilon_\perp})$, and 
the functions $\widetilde{h}_{mn}(k)$ with $m,n=1,2,3$ and $\widetilde{p}(k)$ are the functions 
$h_{mn}(k)$ and $p(k)$ from the Sec.~\ref{sec:trilayer} with the following substitutions:
$r_0\rightarrow \widetilde{r}_0=r_0/\sqrt{\epsilon_\perp}$, 
$L\rightarrow \widetilde{L}_0=L_0/\sqrt{\epsilon_\perp}$, and $\delta\rightarrow \widetilde{\delta}$.   

\subsubsection{Symmetric excitons: type I}
 
In the case of symmetric exciton of the first type (see Fig.~\ref{fig:trilayer_2022_even_1}) the top (t) (or  bottom (b)) charge of conduction band electron quasiparticle is an electron charge, while the charges of the hole quasiparticles are 
\begin{equation}
Q_\mathrm{t}=Q_\mathrm{b}=\frac{|e|}{4}\Big(1+\frac{\Delta_v}{\sqrt{\Delta_v^2+8t^2}}\Big)\approx 0.481|e|,
\end{equation}
for the top and bottom layers, and  
\begin{equation}
Q_\mathrm{m}=\frac{|e|}{2}\Big(1-\frac{\Delta_v}{\sqrt{\Delta_v^2+8t^2}}\Big)\approx 0.038|e|
\end{equation} 
for the charge of the hole, which belongs to the middle (m) layer. Therefore we have the potential
\begin{align}
V_\mathrm{tri}^\mathrm{s,I}(k)=&-\frac{2\pi e^2}{k\sqrt{\epsilon_\perp}}\frac{(1-\widetilde{\delta})}{\widetilde{p}(k)}\times \nonumber \\ 
\times&\big[0.481 \{\widetilde{h}_{11}(k)+\widetilde{h}_{13}(k)\}+0.038 \widetilde{h}_{12}(k)\big].
\end{align}

\subsubsection{Symmetric excitons: type II}

For the case of the second type of symmetric excitons, we have the following distribution of the hole charges 
in the trilayer 
\begin{equation}
Q_\mathrm{t}=Q_\mathrm{b}=\frac{|e|}{4}\Big(1-\frac{\Delta_v}{\sqrt{\Delta_v^2+8t^2}}\Big)\approx 0.019|e|,
\end{equation}
for the top and bottom layers, and  
\begin{equation}
Q_\mathrm{m}=\frac{|e|}{2}\Big(1+\frac{\Delta_v}{\sqrt{\Delta_v^2+4t^2}}\Big)\approx 0.962|e|.
\end{equation} 
The charge of the conduction band electron is localized in the middle layer. Therefore the effective potential is 
\begin{equation}
V_\mathrm{tri}^\mathrm{s,II}(k)=-\frac{2\pi e^2}{k\sqrt{\epsilon_\perp}}\frac{[0.962 \widetilde{h}_{22}(k)+
0.038 (1-\widetilde{\delta}) \widetilde{h}_{12}(k)]}{\widetilde{p}(k)}.
\end{equation}

Presenting all the potentials in the form 
\begin{align}
V_\text{tri}^\alpha(k)=&-\frac{\pi e^2}{\sqrt{\epsilon_\perp}k}U_\text{tri}^\alpha(k),
\end{align}
with $\alpha=$ as; s,~I; s,~II, and introducing the coordinate dependent functions 
\begin{equation}
u_\text{tri}^{\alpha}(\rho)=\int_0^\infty dk J_0(k\rho)U_\text{tri}^{\alpha}(k),
\end{equation}
one can write the eigenvalue equation for each species of the excitons in the form
\begin{equation}
\Big\{\frac{\hbar^2}{2\mu}\nabla_\parallel^2+\frac{e^2}{2\sqrt{\epsilon_\perp}}
u_\text{tri}^\alpha(\rho)+E\Big\}\psi^\alpha(\rho)=0.
\end{equation}
Using the the parameters $\xi=\rho\varepsilon\sqrt{\epsilon_\perp}/r_0$,  
$\epsilon=E/[(\mu e^4/2\hbar^2\varepsilon^2)]$ and the notation 
\begin{equation}
\label{eq:v_tri}
v^\alpha_\text{tri}(\xi)=
\varepsilon \int_0^\infty dx J_0(x\xi)U^\alpha_\text{tri}
\Big(\frac{x\varepsilon\sqrt{\epsilon_\perp}}{r_0}\Big),
\end{equation}
we write the eigenvalue equation for $s$ excitonic states 
\begin{equation}
\Big\{b^2\epsilon_\perp\frac{1}{\xi}\frac{d}{d\xi}\Big(\xi\frac{d}{d\xi}\Big)+
bv_\text{tri}^\alpha(\xi)+\epsilon\Big\}\psi^\alpha(\xi)=0.
\end{equation}

For the case of antisymmetric excitons, we have $l=L\varepsilon/r_0\approx 1.298$, 
where we use the value $L\approx 12.98\mbox{\AA}$
for the thickness of the trilayer.

Solving the eigenvalue equation we obtain $E_1\approx -90$\,meV, $E_2\approx -28$\,meV, 
$E_3\approx -13$\,meV, $E_4\approx -7$\,meV. It gives the corresponding values for $\Delta E_j=E_j-E_1$: 
$\Delta E_2\approx 62$\,meV, $\Delta E_3\approx 77$\,meV, and $\Delta E_3\approx 83$\,meV.

For the symmetrical excitons of the first type we have $E_1\approx -89$\,meV, $E_2\approx-28$\,meV, 
$E_3\approx -13$\,meV, $E_4\approx -7$\,meV. It gives the corresponding values for $\Delta E_j=E_j-E_1$: 
$\Delta E_2\approx 61$\,meV, $\Delta E_3\approx 76$\,meV, and $\Delta E_3\approx 82$\,meV. 

For the symmetrical excitons of the second type we have $E_1\approx -121$\,meV, $E_2\approx-31$\,meV, 
$E_3\approx -15$\,meV, $E_4\approx -7$\,meV. It gives the corresponding values for $\Delta E_j=E_j-E_1$: 
$\Delta E_2\approx 90$\,meV, $\Delta E_3\approx 106$\,meV, and $\Delta E_3\approx 114$\,meV. 

\subsection{Quadrolayer}
\label{subsec:quadro_excitons}

\begin{figure}[t]
	\centering
	\includegraphics[width=0.75\linewidth]{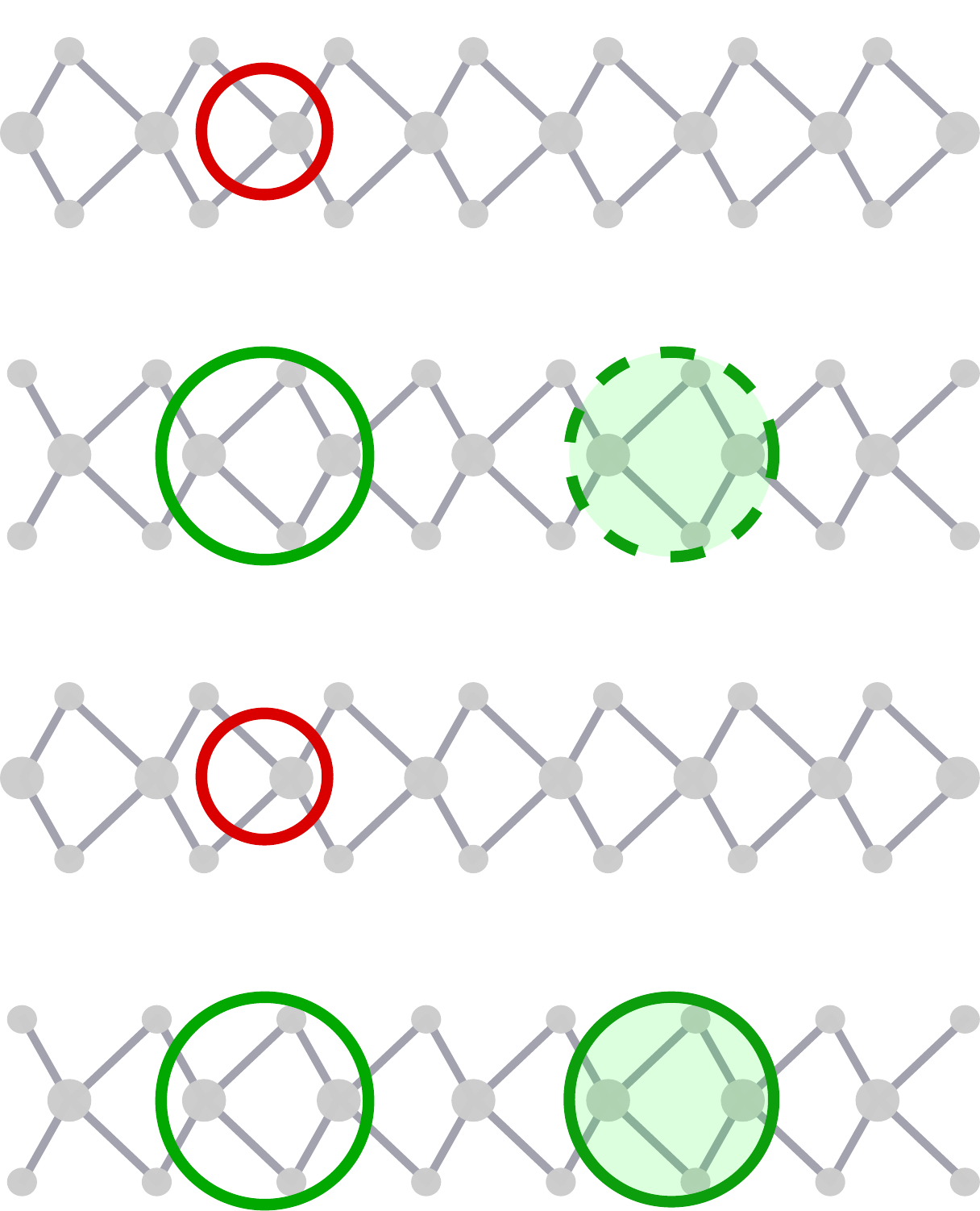}%
	\caption{The schematic picture of the intralayer exciton in the quadrolayer S-TMD in K$^+$ point.  
	The empty circles represent the hole quasiparticle in the trilayer, localized mainly in the first and third layers. 
	The filled circles represent the conduction band electron quasiparticle localized either in the first (solid circle)
	or the third (dashed circle) layer.
	The size of the circles denotes the amount of charge of quasiparticles redistributed in each layer. 
	The colors of the circles encode the polarizations of photons which can be coupled to the exciton: 
	green for $\sigma^+$ and red for $\sigma^-$. The latter means that the corresponding intralayer exciton   
	in K$^+$ point is mainly active in $\sigma^+$ polarization and much smaller in $\sigma^-$ polarization.} 
	\label{fig:quadrolayer}
\end{figure}

The fine structure of the quadrolayer is more complicated than in bi- and trilayer. Following the methodology presented in Ref.~\cite{aSlobodeniuk2019} we first calculate the characteristics of the conduction and valence band 
states. The interlayer coupling between conduction band states is zero in the leading order of $\mathbf{k\cdot p}$-perturbation theory. Therefore, line in the previous cases, one can consider the conduction band excitations localized in 
each layer separately. The valence band excitations are affected by the interlayer interaction. As a result of this interaction, the new valence band states are formed in K$^+$ and K$^-$ valleys of the quadrolayer. We derive these new states in K$^+$ valley for brevity. The states in the opposite valley can be obtained from the previous one by applying the time symmetry operation. Following Ref.~\cite{aSlobodeniuk2019} we introduce the valence band states 
$|\Psi_v^{(j)}\rangle|s\rangle$ in K$^+$ point of quadrolayer, where $j=1,2,3,4$ enumerates the number of the layer, and $s=\uparrow,\downarrow$ is spin index. The Hamiltonian matrix, which describes the interaction between these states, written in the basis $\{|\Psi_v^{(1)}\rangle|s\rangle,|\Psi_v^{(2)}\rangle|s\rangle,|\Psi_v^{(3)}\rangle|s\rangle |\Psi_v^{(4)}\rangle|s\rangle\}$, reads  
\begin{equation}
H^{(4)}_{vs}=\left[
\begin{array}{cccc}
\sigma_s\frac{\Delta_v}{2} & t & 0 & 0 \\
t & -\sigma_s\frac{\Delta_v}{2} & t & 0 \\
0 & t & \sigma_s\frac{\Delta_v}{2} & t \\
0 & 0 & t & -\sigma_s\frac{\Delta_v}{2} \\
\end{array}
\right].
\end{equation}   
Here $\sigma_s=+1/-1$ for $s=\uparrow/\downarrow$, $\Delta_v$ is the spin-splitting in valence bands, 
and $t$ is the interlayer coupling parameter. Taking into account that the quadrolyaer crystal is centrosymmetric
we conclude that the spectrum of valence band excitations is doubly degenerated by spin and, hence, it is enough to consider the case of spin-up state $\sigma_s=1$. Solving the corresponding eigenvalue equation  (see Sec.~\ref{sec:VB_spectrum} for details) we obtain the spectrum of the valence bands excitations
\begin{align}
E_\pm^{(\eta)}= \pm \frac{1}{2}\sqrt{\Delta_v^2+6 t^2+2\eta\sqrt{5} t^2},
\end{align}
where parameter $\eta=\pm 1$ is introduced to separate two different branches of the new valence band excitations. 
The subscript marks the upper (``+'') and lower (``-'') branches of the valence band excitations. We are interested in lowest energy intralayer excitons transitions, which involve only the upper branches. The corresponding eigenstates 
can be written as $|\Phi_+^{(-)}\rangle|\uparrow\rangle$ and $|\Phi_+^{(+)}\rangle|\uparrow\rangle$. 
In the case of the WSe$_2$ quadrolayer with $\Delta_v=456$~meV and $t=67$~meV, we obtain approximately 
the eigenstates of the valence bands 
\begin{widetext}
\begin{align}
|\Phi_{+,v}^{(-)}\rangle \approx&  -0.847|\Psi_v^{(1)}\rangle-0.047|\Psi_v^{(2)}\rangle+
0.524|\Psi_v^{(3)}\rangle+0.076|\Psi_v^{(4)}\rangle, \\
|\Phi_{+,v}^{(+)}\rangle\approx&  +0.513|\Psi_v^{(1)}\rangle+0.187|\Psi_v^{(2)}\rangle+
0.830|\Psi_v^{(3)}\rangle+0.116|\Psi_v^{(4)}\rangle,
\end{align}  
\end{widetext}
with the energies $E_+^{(-)}\approx 232$\,meV and $E_+^{(+)}\approx 252$\,meV, respectively.  
The structure of these composite hole states is presented in Fig.~\ref{fig:quadrolayer}. 
The decomposition coefficients $q_j^{(\pm)}$ of the states $|\Phi_{+,v}^{(\pm)}$ by the basis states 
$|\Psi_v^{(j)}\rangle$ define the distribution of the charges $Q_j^{(\pm)}=[q_j^{(\pm)}]^2|e|$ 
in the corresponding hole states. 

As one can see, for both cases the charge of the holes is placed mainly in the first and the third layers. Namely 
for $|\Phi_+^{(-)}\rangle$ they are $Q_1^{(-)}\approx 0.72|e|$ and $Q_3^{(-)}\approx 0.27|e|$, and  
for $|\Phi_+^{(+)}\rangle$ they are $Q_1^{(+)}\approx 0.26|e|$ and $Q_3^{(+)}\approx 0.69|e|$.
In other words, 99\% and 95\% of the total charge of the hole are localized in the first and third layers, 
for each case respectively. On the contrary, the charge of the conduction band electron is fully localized 
only within one of the layers. There are two possibilities for the localization of the conduction electrons
which are related to the studied intralayer exciton transitions.  

Hence, due to optical transition the charge from $|\Phi_{+,v}^{(\pm)}\rangle|\uparrow\rangle$ state transfers 
to the either $|\Psi_{c}^{(1)}\rangle|\uparrow\rangle$ or to the state $|\Psi_{c}^{(3)}\rangle|\uparrow\rangle$. 
Here $|\Psi_{c}^{(j)}\rangle|\uparrow\rangle$ is the conduction electron Bloch spin-up state in the K$^+$ point 
of the $j$-th layer. 

Therefore, 4 types of transitions correspond to the lowest-energy intralayer excitons in the quadrolayer. 

The first type of transitions $|\Phi_{+,v}^{(-)}\rangle\rightarrow |\Psi_c^{(1)}\rangle$, 
is characterized by the optical strength $f_1$. It is proportional to the square of the matrix element    
\begin{equation}
f_1\propto |\langle\Psi_c^{(1)}|\widehat{\mathbf{p}}|\Phi_{+,v}^{(-)}\rangle|^2.  
\end{equation}
Here $\widehat{\mathbf{p}}$ is an electron momentum operator. Taking into account the structure of the valence
band state we obtain $f_1\approx 0.72f$, where $f$ is the optical strength of the excitonic transition in monolayer. 
For the second type of the transitions  $|\Phi_{+,v}^{(+)}\rangle\rightarrow |\Psi_c^{(1)}\rangle$, 
we have $f_2\approx 0.26f$. 
For the third type of transition $|\Phi_{+,v}^{(-)}\rangle\rightarrow |\Psi_c^{(3)}\rangle$, we have  
$f_3\approx 0.27f$. 
For the fourth type of transitions 
$|\Phi_{+,v}^{(+)}\rangle\rightarrow |\Psi_c^{(3)}\rangle$, we have $f_4\approx 0.69f$.   

Using this approximation one can find the Coulomb interaction which is responsible 
for the formation of the excitons, with the electron localized in the first layer 
\begin{align}
V_1^{(\pm)}(k)=-\frac{2\pi e^2}{k\sqrt{\epsilon_\perp}}\frac{(1-\widetilde{\delta})}{\widetilde{g}(k)}
\sum_{j=1}^4\widetilde{f}_{1j}(k)[q_j^{(\pm)}]^2,
\end{align} 
and in the third layer 
\begin{widetext}
\begin{align}
V_3^{(\pm)}(k)=-\frac{2\pi e^2}{k\sqrt{\epsilon_\perp}}\frac{\widetilde{f}_{22}(k)[q_3^{(\pm)}]^2+
\widetilde{f}_{23}(k)[q_2^{(\pm)}]^2+(1-\widetilde{\delta})\widetilde{f}_{12}(k)[q_4^{(\pm)}]^2+
(1-\widetilde{\delta})\widetilde{f}_{13}(k)[q_1^{(\pm)}]^2}{\widetilde{g}(k)}.
\end{align}  
\end{widetext} 
Here $\widetilde{\delta}=(\varepsilon-\sqrt{\epsilon_\perp})/(\varepsilon-\sqrt{\epsilon_\perp})$, and 
the functions $\widetilde{f}_{mn}(k)$ with $m,n=1,2,3,4$ and $\widetilde{g}(k)$ are the functions 
$f_{mn}(k)$ and $g(k)$ from the Sec.~\ref{sec:quadrolayer} with the following substitutions:
$r_0\rightarrow \widetilde{r}_0=r_0/\sqrt{\epsilon_\perp}$, 
$L\rightarrow \widetilde{L}_0=L_0/\sqrt{\epsilon_\perp}$, and 
$\delta\rightarrow \widetilde{\delta}=(\delta-\sqrt{\epsilon}_\perp)/(\delta+\sqrt{\epsilon}_\perp)$.

Using the notation for both potentials $V_\alpha^{(\pm)}(k)=(-2\pi e^2/k\sqrt{\epsilon_\perp})U^{(\pm)}_\alpha(k)$, with $\alpha=1,3$, and introducing the function 
\begin{equation}
u^{(\pm)}_\alpha(\rho)=\int_0^\infty dk U^{(\pm)}_\alpha(k)J_0(k\rho),
\end{equation}
we write the eigenvalue equation for the spectrum of the excitons in the form 
\begin{equation}
\Big\{\frac{\hbar^2}{2\mu}\nabla_\parallel^2+\frac{e^2}{\sqrt{\epsilon_\perp}}
u^{(\pm)}_\alpha(\rho)+E\Big\}\psi^{(\pm)}_\alpha(\rho)=0.
\end{equation}
Introducing again the dimensionless parameters of energy $\epsilon=E/(\mu e^4/2\hbar^2\varepsilon^2)$ 
and distance $\xi=\rho\varepsilon/\widetilde{r_0}$ we obtain the following dimensionless equation
for $ns$ excitonic states 
\begin{equation}
\Big\{b^2\epsilon_\perp\frac{1}{\xi}\frac{d}{d\xi}\Big(\xi\frac{d}{d\xi}\Big)+
bv_\alpha^{(\pm)}(\xi)+\epsilon\Big\}\psi^{(\pm)}_\alpha(\xi)=0,
\end{equation} 
where 
\begin{equation}
\label{eq:v_quadro}
v^{(\pm)}_\alpha(\rho)=2\varepsilon\int_0^\infty dx J_0(x\xi)U^{(\pm)}_\alpha
\Big(\frac{x\varepsilon\sqrt{\epsilon_\perp}}{r_0}\Big).
\end{equation}

We consider first the situation, where the electron state of the exciton is localized in the first (bottom) 
monolayer's plane. Taking into account that $L=25.96$\,\mbox{\AA}. 

The spectrum for $|\Phi_{+,v}^{(-)}\rangle\rightarrow |\Psi_c^{(1)}\rangle$ is 
$E_1\approx -104$\,meV, $E_2\approx -28$\,meV, $E_3\approx -13$\,meV, $E_4\approx -7$\,meV. 
We have $\Delta E_2\approx 76$\,meV, $\Delta E_3\approx 91$\,meV, $\Delta E_4\approx 97$\,meV. 

The spectrum for $|\Phi_{+,v}^{(+)}\rangle\rightarrow |\Psi_c^{(1)}\rangle$ is 
$E_1\approx -69$\,meV, $E_2\approx -24$\,meV, $E_3\approx -12$\,meV, $E_4\approx -6$\,meV. 
We have $\Delta E_2\approx 45$\,meV, $\Delta E_3\approx 57$\,meV, $\Delta E_4\approx 63$\,meV. 

The spectrum for $|\Phi_{+,v}^{(-)}\rangle\rightarrow |\Psi_c^{(3)}\rangle$ is 
$E_1\approx -67$\,meV, $E_2\approx -24$\,meV, $E_3\approx -12$\,meV, $E_4\approx -6$\,meV. 
We have $\Delta E_2\approx 43$\,meV, $\Delta E_3\approx 55$\,meV, $\Delta E_4\approx 61$\,meV. 

The spectrum for $|\Phi_{+,v}^{(+)}\rangle\rightarrow |\Psi_c^{(3)}\rangle$ is 
$E_1\approx -100$\,meV, $E_2\approx -27$\,meV, $E_3\approx -13$\,meV, $E_4\approx -7$\,meV. 
We have $\Delta E_2\approx 73$\,meV, $\Delta E_3\approx 87$\,meV, $\Delta E_4\approx 93$\,meV. 

Note that there are two groups of exciton series. The first group has the binding energy $E_1\approx -100$\,meV
of the ground state exciton, while the second group has the binding energy $E_1\approx -70$\,meV.

We summarized all the results for the spectrum of the excitons in the multilayers in the Table~\ref{tab:differencies}
\begin{widetext}
\begin{center}
\begin{table}[h]
 %\begin{center}
 \begin{tabular}{p{2cm} p{1.5cm}  p{1.5cm}  p{1.5cm} p{1.5cm} p{1.5cm}  p{1.5cm} p{1.5cm} p{1.5cm} p{1.0cm}}
 \hline\hline \\ [-0.5ex]
 $\Delta E_n$\,[meV] & ML & BL  & TL$^{(1)}$ & TL$^{(2)}$ & TL$^{(3)}$ & QL$^{(1)}$ & QL$^{(2)}$ & QL$^{(3)}$ & QL$^{(4)}$  \\ [1.0ex]
 \hline \\
 $\Delta E_2$  & 126  & 105 & 62 & 61 & 90  &  76  &  45 &  43 &  73 \\ [1.5ex]
 $\Delta E_3$  & 148  & 124 & 77 & 76 & 106 &  91  &  57 &  55 &  87 \\ [1.5ex]
 $\Delta E_4$  & 156  & 131 & 83 & 82 & 114 &  97  &  63 &  61 &  93 \\ [1.5ex]
 \hline
\end{tabular}
%\end{center}
\caption{\label{tab:differencies}
Differences between the binding energies $(E_n)$ of $n(=2,3,4)$th excited and of the ground $(E_1)$ 
intralayer exciton states $\Delta E_n=E_j-E_1$ in the mono-(ML), bi-(BL), tri-(TL), and quadlayer (QL).
The signs TL$^{(1)}$, TL$^{(2)}$, and TL$^{(3)}$ represent the ``as'', ``sI'' and ``sII'' excitons in trilayer, respectively. 
The signs QL$^{(1)}$, QL$^{(2)}$, QL$^{(3)}$, and QL$^{(4)}$ represent the excitons in quadlayer, which corresponds to 
$|\Phi_{+,v}^{(-)}\rangle\rightarrow |\Psi_c^{(1)}\rangle$, $|\Phi_{+,v}^{(+)}\rangle\rightarrow |\Psi_c^{(1)}\rangle$, 
$|\Phi_{+,v}^{(-)}\rangle\rightarrow |\Psi_c^{(3)}\rangle$, and $|\Phi_{+,v}^{(+)}\rangle\rightarrow |\Psi_c^{(3)}\rangle$ transitions, 
respectively.}
\end{table}
\end{center}  
\end{widetext}

We used the calculated eigenfunctions of the excitons in the multilayers to estimate the diamagnetic shifts $\alpha_n^{(N)}$
of the corresponding excitons. The results of this estimation are presented on Table~\ref{tab:diam_shifts}.
\begin{widetext}
\begin{center}
\begin{table}[h]
 %\begin{center}
 \begin{tabular}{p{2.5cm} p{1.5cm}  p{1.5cm}  p{1.5cm} p{1.5cm} p{1.5cm}  p{1.5cm} p{1.5cm} p{1.5cm} p{1.0cm}}
 \hline\hline \\ [-0.5ex]
 $\alpha_n^{(N)}$\,[$\mu$eV$\cdot$T$^{-2}$] & ML & BL  & TL$^{(1)}$ & TL$^{(2)}$ & TL$^{(3)}$ & QL$^{(1)}$ & QL$^{(2)}$ & QL$^{(3)}$ & QL$^{(4)}$  \\ [1.0ex]
 \hline \\
 $1s$  & 0.28  & 0.32 & 0.6  & 0.6  & 0.36  &  0.45  &  0.9  &  0.9  &  0.5 \\ [1.5ex]
 $2s$  & 4.7   & 5.6  & 8.2  & 8.2  & 6.5   &  7.6   &  10.5 &  10.6 &  7.9 \\ [1.5ex]
 $3s$  & 25    & 28   & 36.5 & 36.5 & 31    &  35    &  44   &  44   &  36 \\  [1.5ex]
 $4s$  & 79    & 87   & 107  & 107  & 95    &  105   &  123  &  124  &  107 \\ [1.5ex]
 \hline
\end{tabular}
%\end{center}
\caption{\label{tab:diam_shifts}
Diamagnetic shifts $\alpha_n^{(N)}$ of $n(=2,3,4)$th excited and of the ground $n=1$ 
intralayer exciton states  in the mono-(ML, $N=1$), bi-(BL, $N=2$), tri-(TL,$N=3$), and quadlayer (QL,$N=4$). }
\end{table}
\end{center}  
\end{widetext}

\end{document}